\documentclass[fleqn,usenatbib]{mnras}

\usepackage{newtxtext,newtxmath}
\usepackage{hyperref}
\usepackage[T1]{fontenc}

\pdfoutput=1

\usepackage{graphicx}	
\usepackage{amsmath}	
\usepackage{amssymb}	

\title[SMGs and top-heavy IMF]{Sub-millimeter galaxies in hierarchical models: revisiting the need for a top-heavy stellar initial mass function with Bayesian optimisation}

\author[Edward J. Elliott et al.]{
Edward J. Elliott,$^{1,2,3}$
Carlton M. Baugh,$^{1,2}$\thanks{E-mail: c.m.baugh@durham.ac.uk}
Cedric G. Lacey,$^{1}$
\\
$^{1}$Institute for Computational Cosmology, Department of Physics, Science Laboratories, South Road, Durham, DH1 3LE, UK\\
$^{2}$Institute for Data Science, Science Laboratories, South Road, Durham, DH1 3LE, UK\\
$^{3}$Engelhart, 7th Floor, Berkeley Square House, 4-19 Berkeley Square, London, W1J 6BR, UK
}

\date{Accepted XXX. Received YYY; in original form ZZZ}

\pubyear{2024}

\begin{document}
\label{firstpage}
\pagerange{\pageref{firstpage}--\pageref{lastpage}}
\maketitle

\begin{abstract}
The properties of high-redshift sub-millimetre galaxies (SMGs) remain controversial within hierarchical structure formation models. We revisit whether a top-heavy stellar initial mass function (IMF) in starbursts is required to reproduce both SMG observations and local galaxy properties. Using Bayesian optimisation, we perform an extensive search of the 15-dimensional parameter space of the \texttt{GALFORM} semi-analytical model. This efficient approach converges to optimal parameter values in fewer than 200 model evaluations, representing orders of magnitude fewer runs than traditional methods.
We test whether \texttt{GALFORM} can simultaneously match three key observational constraints: the $z=0$ $K$-band luminosity function, the SMG number counts at 850~$\mu$m, and the SMG redshift distribution. We consider two model variants: one with a universal solar neighbourhood IMF for all star formation, and another allowing the IMF slope in starbursts to vary as a free parameter. When assuming a universal Chabrier IMF, we find no parameter combination that simultaneously reproduces all three datasets. The model either matches the SMG constraints while grossly overpredicting the local $K$-band luminosity function, or matches the local luminosity function while severely underpredicting SMG counts by factors of 3--100.
In contrast, allowing a top-heavy IMF in starbursts enables excellent simultaneous fits to all constraints. The best-fitting model prefers an IMF slope parameter $x \approx 0.7$ (where d$n$/dlog$m \propto m^{-x}$), somewhat more top-heavy than recent models but less extreme than early proposals. Our comprehensive parameter space exploration definitively confirms that, within the \texttt{GALFORM} framework, a top-heavy IMF in starbursts is necessary to reconcile high-redshift dusty star-forming galaxies with local galaxy populations.
\end{abstract}

\begin{keywords}
methods:statistical -- methods:numerical -- galaxies:formation
\end{keywords}



\section{Introduction}
Observations of high-redshift galaxies provide powerful constraints on galaxy formation models, particularly when combined with measurements of local galaxy populations. Sub-millimetre galaxies (SMGs), first discovered using the SCUBA instrument on the James Clerk Maxwell Telescope \citep{Smail1997,Hughes1998}, represent a population undergoing intense dust-obscured star formation at high redshift \citep[for a review, see][]{Casey2014}. The sub-millimetre emission from these galaxies benefits from a negative $k$-correction, such that their observed flux remains nearly constant over a wide range of redshifts for a fixed total infrared luminosity. This property provides a unique window for studying galaxy evolution across cosmic time.

Bright SMGs contribute up to half the star formation rate density at $z \sim 2$--3 despite their low space density of $\sim 10^{-5}$~cMpc$^{-3}$ \citep{Chapman2005,Smith2017}. They possess large stellar masses of $\sim 10^{11}$~M$_\odot$ \citep{Swinbank2004,DaCunha2015} and reside in massive dark matter halos of $\sim 10^{13}$~M$_\odot$ \citep{Blain2004,Hickox2012}. Assuming their emission is powered by star formation with a solar neighbourhood IMF yields intense star formation rates, suggesting that the SMG phase could build a substantial fraction of the stellar mass in present-day massive elliptical galaxies.

A viable galaxy formation model should reproduce both the number counts and redshift distribution of SMGs while simultaneously matching observations of local galaxies. This has proven remarkably challenging. Early semi-analytic models underestimated SMG number counts by more than an order of magnitude \citep{Granato2000,Somerville2012}. \citet{Baugh2005} proposed that adopting a stellar initial mass function in starbursts with a larger proportion of massive stars than in the solar neighbourhood (specifically, d$n$/dlog$m \propto m^{-x}$ with $x = 0$; compared to $x = 1.35$ above one solar mass for a Chabrier or Salpeter IMF) allowed the model to match both SMG number counts and redshift distributions without compromising low-redshift predictions. In this model, both dust absorption of starlight and the spectral energy distribution of dust emission were calculated self-consistently using a physical model that included the dependence of dust temperature on dust mass. Later model iterations \citep{Lacey2016,Baugh2019,Cowley2019}, which include AGN heating feedback to regulate the abundance of bright galaxies, reproduce SMG and local galaxy properties with a somewhat less extreme but still top-heavy IMF in bursts, with $x \sim 1$.

\citet{McAlpine2019} presented predictions from the EAGLE hydrodynamic simulation \citep{Schaye2015}, post-processed with a dust radiative transfer code, assuming a Chabrier IMF for all star formation. These authors found some agreement between the model and the SMG redshift distribution for $z > 1$, but the simulation greatly underpredicted number counts, with essentially no sources brighter than 5~mJy \citep[see comparison in][]{Cowley2019}. The SIMBA hydrodynamic simulation \citep{Dave2019,Lovell2021}, also assuming a Chabrier IMF, similarly produces reasonable agreement with the SMG redshift distribution but underpredicts counts by factors of 3--10 at the bright end (depending on assumptions about source blending) and by more than an order of magnitude at the faint end. \citet{Hayward2021} compared sub-millimetre predictions for the Illustris and IllustrisTNG simulations \citep{Nelson2015,Pillepich2018} using the fitting formula from \citet{Hayward2011}. Illustris reproduces number counts reasonably well but predicts a significantly lower median redshift for bright sources than observed, while IllustrisTNG produces a better redshift distribution but underpredicts SMG number counts by more than an order of magnitude at bright fluxes.

Here we revisit the need for a top-heavy IMF in hierarchical galaxy formation models. Allowing the IMF to vary with star formation mode has been controversial \citep[see, for example,][]{Bastian2010}. Nevertheless, observational evidence for IMF variations exists. Using emission line strengths and Balmer decrements from star-forming galaxies in the Galaxy and Mass Assembly Survey, \citet{Gunawardhana2011} argued that the IMF slope varies with star formation rate, becoming more top-heavy as the rate increases, reaching $x \approx 0.9$ in the most intense star-forming galaxies---similar to the value adopted by \citet{Lacey2016}. \citet{Romano2017} inferred a top-heavy IMF with $x \approx 0.95$ in nearby starburst galaxies based on CNO isotopic ratios in the ISM, extended by \citet{Zhang2018} to starbursts at $z \sim 2$--3. \citet{Schneider2018a}, studying massive stars in the Large Magellanic Cloud, found $x = 0.9 \pm 0.3$, with subsequent analysis by \citet{Farr2018} finding $x = 1.05 \pm 0.14$, both top-heavy compared to the solar neighbourhood IMF. While many studies provide evidence for IMF variations, the exact nature remains uncertain. Some studies \citep[e.g.,][]{Conroy2012,Smith2020} find evidence for bottom-heavy IMFs in high-mass galaxies (noting that an IMF can be simultaneously both top-heavy and bottom-heavy compared to a Chabrier IMF), while \citet{Weidner2013} propose a time-dependent IMF favouring massive stars at early times and low-mass stars at late times.

All galaxy formation models---whether hydrodynamic simulations or semi-analytic codes---involve numerous parameters governing sub-grid processes such as star formation and feedback from supernovae and active galactic nuclei \citep{Baugh2006,Benson2010a,Crain2015,Somerville2015}. These parameters must be calibrated to match some subset of observations before predictions can be made for other observables. One criticism of including a top-heavy IMF in \texttt{GALFORM} is that, given the relatively large parameter space, another unexplored parameter combination might fit observations without this assumption. We aim to robustly determine whether a parameter set exists that allows the model, assuming a universal solar neighbourhood IMF, to match SMG observational constraints while maintaining reasonable fits to low-redshift datasets such as the $K$-band luminosity function. As a secondary aim, we investigate the required IMF slope in starbursts: will a more extensive parameter space search reveal the possibility of a less extreme IMF in bursts?

To achieve these aims, we employ Bayesian optimisation techniques \citep[see e.g.][]{Frazier2018}. This approach calibrates model parameters by searching for the best-fitting parameter set, judged by how well the model reproduces target datasets. The fitting metric measures the distance between model predictions and calibration data. This represents a ``black-box'' problem: we do not know how the metric depends on model parameters. Bayesian optimisation is designed for computationally expensive models requiring optimisation methods that use few model evaluations. In Bayesian optimisation, the metric is estimated at every point in parameter space with varying uncertainty. We use Gaussian processes \citep[see e.g.][]{GP} to describe the metric. As more model evaluations are performed, the Gaussian process is updated. Once no further significant improvement occurs in the metric, the best-fitting model has the lowest metric value.

We show that this method can find suitable fits in a small fraction of the model evaluations required by other, often more elaborate, approaches \citep[see e.g.][]{Kampakoglou2008,Henriques2009,Bower2010,Vernon2010,Benson2010b,Lu2011,Lu2012,Ruiz2015,Martindale2017,vanderVelden2019,vanderVelden2021,Elliott,Makun}. However, this approach provides limited information about uncertainty in the best-fit parameters in exchange for requiring an order of magnitude fewer full model evaluations. Bayesian optimisation has been successfully applied to other computationally expensive astronomical problems involving high-dimensional parameter spaces, including interpreting supernovae light curves \citep{Leclercq2018}, analysing the Lyman-$\alpha$ forest power spectrum \citep{Rogers2019,Takhtaganov2021}, and searching for signatures of inflation in cosmic microwave background data \citep{Hamann2022}.

Semi-analytical model calibration and parameter space exploration has generally taken two forms: (1) direct exploration using full model calculations, and (2) emulation, where a small number of full model runs are performed with the bulk of parameter sampling done by a computationally cheap surrogate. Although semi-analytical models are much cheaper than hydrodynamic simulations, direct parameter space exploration remains computationally expensive due to the large number of model runs required for formal searches of high-dimensional spaces.

Examples of the first approach using full model runs include \citet{Kampakoglou2008}, who implemented Markov Chain Monte Carlo (MCMC) techniques to calibrate a semi-analytic model to multiple observational datasets. \citet{Henriques2009} used MCMC to calibrate the L-GALAXIES model to several datasets, finding that the choice of datasets altered the values of the best-fitting parameters, pointing to model deficiencies. \citet{Lu2011,Lu2012} constrained the parameter space providing acceptable fits to the $K$-band luminosity function, expanding this to include the HI mass function in \citet{Lu2014}. \citet{Ruiz2015} used particle swarm optimisation to calibrate a semi-analytic model to the $K$-band luminosity function.

The second class of methods involves constructing a statistical emulator of the semi-analytic model that can be evaluated orders of magnitude faster than running the model itself, at the cost of being approximate. \citet{Bower2010} and \citet{Vernon2010} employed a Bayesian emulation technique \citep[developed by][]{bayeslinear} to constrain the \texttt{GALFORM} parameter space region providing reasonable fits to the $K$- and $b_J$-band luminosity functions \citep[see also][]{Benson2010b}. The same approach was applied by \citet{Rodrigues2017} to calibrate \texttt{GALFORM} to the stellar mass function in the local Universe and its evolution, and by \citet{vanderVelden2021} to calibrate the Meraxes semi-analytic model at high redshift. The Bayes linear approach used by \citet{Bower2010} makes approximations about the functions being minimised and is not strictly a black-box method like Bayesian optimisation. The \citet{Bower2010} method involved searching the parameter space in waves, with the space redefined at each wave to make the search adaptive. \citet{Bower2010} used 5500 runs of GALFORM in their search of a parameter space similar in size to that considered here; we use fewer than 200 runs of the full model in one sweep of the parameter space, making the search fully automatic.

We presented a new framework for automated GALFORM calibration using local observations in \citet{Elliott}. The first step was sensitivity analysis to determine which model parameters most strongly shaped predictions for the calibration data (see \citealt{Oleskiewicz2020} for the first application of sensitivity analysis to \texttt{GALFORM}). This identified a subset of 10 parameters as most important for determining the calibration data. We ran \texttt{GALFORM} for 1000 parameter sets sampled from this 10-dimensional space using a Latin hypercube \citep{Stein1987}. These runs allowed us to build an emulator of \texttt{GALFORM} using an artificial neural network. The emulator enabled extensive MCMC searches of the parameter space, returning best-fitting parameter sets for chosen calibration data along with the range of acceptable model predictions. The speed of this approach allowed us to explore best-fitting models resulting from different calibration dataset combinations. Differences in resulting best-fitting models point to possible model deficiencies or to incompatibilities between observational measurements. \cite{Makun} applied a similar framework to the calibration of \texttt{GALFORM} to reproduce the number counts of $H_{\alpha}$ emitters.

Here we face a more challenging problem. First, the parameter space is larger than in \citet{Elliott}, with 15 dimensions instead of 10. The focus of the parameter space search is now: ``Can we find any example of a model that works under these assumptions?'' rather than finding a best-fitting parameter set with its associated range of acceptable models. Second, the computational overhead for each parameter set is much higher. We need predictions for galaxy number counts and redshift distributions, which require running \texttt{GALFORM} at many redshifts rather than just $z=0$. Moreover, some predictions are sensitive to rare events such as starbursts, requiring many more dark matter halo merger histories to obtain robust predictions.

To overcome these challenges, we investigate a new approach to model calibration using Bayesian optimisation. This global optimisation technique efficiently searches the parameter space using a Gaussian process prior and is capable of efficiently searching high-dimensional parameter spaces for global minima. We use this method to test whether a parameter set exists that can match SMG observations and low-redshift constraints simultaneously without including a top-heavy IMF. Past parameter space explorations, though usually performed manually, have suggested this is not possible when assuming a universal solar neighbourhood IMF. We aim to test this conclusion with a more sophisticated parameter search over a comprehensive list of relevant parameters and a wide search space. This approach represents an enormous improvement over the old-fashioned one-parameter-at-a-time searching originally used to argue for a top-heavy IMF.

This paper is structured as follows: in \S2, we briefly introduce \texttt{GALFORM}, focusing on processes and parameters relevant to our study. In \S3, we review the theory and practical considerations behind Bayesian optimisation. We list the observational datasets used for calibration in \S3.5 and validate our method on a surrogate model in \S3.6. In \S4, we present the model calibration for different assumptions about the IMF and about the importance placed on reproducing various datasets. We give our conclusions in \S5.


\section{Galaxy formation model}
\label{sec:maths} 

We use the \texttt{GALFORM} semi-analytical model of galaxy formation introduced by \cite{Cole:2000} (see also \citealt{Bower2006} and \citealt{Lacey2016}). \texttt{GALFORM} models the key physical process thought to shape the formation and evolution of galaxies in the cold dark matter cosmology (for reviews of the semi-analytical approach to modelling galaxy formation see \citealt{Baugh2006} and \citealt{Benson2010a}). The model starts from a sample of dark matter halos along with their formation and merger histories, which is typically extracted from a cosmological simulation of the dark matter evolution. The model then tracks the transfer of mass and metals between different reservoirs of baryons, with each halo typically containing both a central galaxy and satellite galaxies. It predicts the full star formation history of each galaxy in a halo, along with the chemical evolution of the gas.  Two modes of star formation are considered: quiescent star formation in galaxy disks and bursts of star formation triggered by galaxy mergers or by bar instabilities in disks. \texttt{GALFORM} combines this information with a stellar population synthesis model to predict the luminosities of galaxies in different bands (see \citealt{Gonzalez-Perez2014}). Here we use the model implemented in the P-Millennium cosmological $N$-body simulation of the evolution of the dark matter distribution \citep{Baugh2019}. 

 Below we introduce the parameters that are varied in the calibration and explain how they affect various processes. For a full list of the model parameters and a complete description of the model, see \cite{Lacey2016}. The list of model parameters varied, and the ranges considered for each parameter are given in Table~\ref{tab:ParamRanges}.

\subsection{Quiescent star formation in disks}

\texttt{GALFORM} uses an empirical star formation law formulated by \cite{Blitz2006} based on observations of local star-forming disk galaxies, which was implemented by \cite{Lagos2011}. The SFR in the disc is given by 
\begin{equation}
    \psi\textsubscript{disk} = \nu\textsubscript{SF}M\textsubscript{mol, disk}, 
\end{equation}
where $M\textsubscript{mol, disk}$ is the mass of molecular gas in the disk, and $\nu\textsubscript{SF}$ is a constant which we treat as an adjustable parameter within a range suggested by the observations \citep{Bigiel2011}. The fraction of cold gas in the molecular phase depends on the gas pressure in the mid-plane of the disc (see \citealt{Lagos2011}).

\subsection{Galaxy mergers}

In the model, galaxy mergers can trigger bursts of star formation and destroy galactic disks. Mergers occur between satellite and central galaxies, on a timescale controlled by dynamical friction on the satellite due to the main halo. We calculate this merger timescale using the improved galaxy merger model of \cite{Simha2017}, in which satellite subhalos are tracked in the dark matter simulation, up until a subhalo is longer resolved, after which the remaining time for the satellite to merge is calculated analytically based on dynamical friction arguments (see \citealt{Campbell2015} for details of the implementation of this model in \texttt{GALFORM}). We define two  thresholds, $f\textsubscript{ellip}$ and $f\textsubscript{burst}$, which determine the outcome when a satellite galaxy with baryonic mass $M\textsubscript{b, sat}$ merges with a central galaxy with baryonic mass $M\textsubscript{b, cen}$. First, if $M\textsubscript{b, sat}/M\textsubscript{b, cen} \geqslant f\textsubscript{ellip}$ the merger is classified as a \textit{major} merger, and the disk component of the galaxy is destroyed and is added to the spheroid component of the galaxy if one exists, or a new spheroid is formed. The cold gas in the disk is assumed to be consumed in a burst of star formation and is added to the spheroid. Second, if $M\textsubscript{b, sat}/M\textsubscript{b, cen} < f\textsubscript{ellip}$, the merger is classified as \textit{minor}, and the disk survives. In this case, the cold gas in the disk is consumed in a starburst which adds stars to the spheroid if a second condition is met, namely if  $M\textsubscript{b, sat}/M\textsubscript{b, cen} \geqslant f\textsubscript{burst}$.  $f\textsubscript{burst}$ and $f\textsubscript{ellip}$ are free parameters. 

\begin{table}
	\centering
	\caption{The parameters explored in this work and the range of values over which they are varied. The first column gives the symbol for the parameter (and units where relevant), the second gives the process or effect the parameter is involved in, and the third column gives the range of parameter values probed. }
	\label{tab:ParamRanges}
	\begin{tabular}{lccr} 
		\hline
		Name & Process & Range \\
		\hline
  		$\nu\textsubscript{SF}$ (Gyr$^{-1}$) & Quiescent star formation & 0.1 - 4.0\\
		$f\textsubscript{ellip}$ & Galaxy mergers & 0.2 - 0.5\\
		$f\textsubscript{burst}$ & Galaxy mergers & 0.01 - 0.3\\
        $F\textsubscript{stab}$ & Disk instability & 0.5 - 1.2\\
 		$\tau\textsubscript{*burst, min}$ (Gyr) & Burst star formation & 0.01 - 0.05\\

	    $x$ & IMF slope in bursts & 0 - 1.35\\

  		$V\textsubscript{SN, disk}$ (km\,s$^{-1}$) & SN feedback & 10 - 800\\
		$V\textsubscript{SN, burst}$ (km\,s$^{-1}$) & SN feedback & 10 - 800\\

		$\gamma\textsubscript{SN}$ & SN feedback & 1.0 - 4.0\\
		$\alpha\textsubscript{reheat}$ & SN feedback & 0.2 - 3.0\\

        $f\textsubscript{SMBH}$ & BH growth & 0.001 - 0.05\\
		$\alpha\textsubscript{cool}$ & AGN feedback & 0.2 - 4.0\\

 	      $f\textsubscript{cloud}$ & Dust & 0.2 - 0.8\\
		$t\textsubscript{esc}$ (Gyr) & Dust & 0.0001 - 0.01\\
		$\beta\textsubscript{burst}$ & Dust & 1.5 - 1.2\\

		\hline
	\end{tabular}
\end{table}

\subsection{Disk instabilities}

Galactic disks that are dominated by rotational motion can become unstable to bar formation if their degree of self-gravity is too large. The \texttt{GALFORM} model follows \citet{Efstathiou1982}, and assumes that disks become unstable if the following criterion is met:  
\begin{equation}
    \label{eq:diskinstab}
    \frac{V\textsubscript{c}(r\textsubscript{disk})}{(1.68\, GM\textsubscript{disk}/r\textsubscript{disk})^{1/2}} \le F\textsubscript{stab}, 
\end{equation}
where $M\textsubscript{disk}$ is the total disk mass and $r\textsubscript{disk}$ is the disk half-mass radius. Numerical simulations of exponential stellar disks by \citet{Efstathiou1982} found a value of $F\textsubscript{stab} \approx 1.1$. while \citet{christodoulou1995} found a value of 0.9 for gaseous disks. A value of 0.61 or below corresponds to universally stable disks, since this is the value of the left hand side of Eqn.~\ref{eq:diskinstab} for a completely self-gravitating disk. We allow the parameter $F\textsubscript{stab}$ to vary within a reasonable range (see Table~\ref{tab:ParamRanges}). We assume that disks that become unstable are disrupted by bar instabilities on a sub-resolution timescale, such that all the mass is instantaneously transferred to the spheroid and any gas present takes part in a burst of star formation, adding stars to the spheroid.

\subsection{Starbursts}

Bursts of star formation, triggered by mergers or bar instabilities in dynamically unstable disks, are assumed to form stars at a rate
\begin{equation}
    \psi\textsubscript{burst} = \nu\textsubscript{SF, burst}M\textsubscript{cold,burst} = \frac{M\textsubscript{cold,burst}}{\tau\textsubscript{*burst}},
\end{equation}
where the star formation timescale is given by 
\begin{equation}
    \tau\textsubscript{*burst} = \text{max}[f\textsubscript{dyn}\tau\textsubscript{dyn,bulge},\tau\textsubscript{*burst,min}].
\end{equation}
Here the bulge dynamical time is defined as $\tau\textsubscript{dyn, bulge}$ = $r\textsubscript{bulge}/V\textsubscript{c}(r\textsubscript{bulge})$, where the velocity is the circular velocity at the half-mass radius of the bulge. The minimum timescale of a burst, $\tau\textsubscript{*burst,min}$, is treated as an adjustable parameter in the range 1-100Myr. $f_{\rm dyn}$ is held at the value of 20 used by \citet{Lacey2016}.

\subsection{Supernova feedback}

Supernova explosions lead to gas being ejected from galaxies and out of the halo. The model assumes that the mass ejection rate is proportional to the instantaneous star formation rate, $\psi$, with a mass loading factor dependent on the circular velocity of the galaxy, $V_{\rm c}$:
\begin{equation}
    \label{eq:SNfeedback}
    \dot{M}_{\rm{eject}}= \left(\frac{V_{\rm{c}}}{V_{\rm{SN}}}\right)^{-\gamma_{\rm{SN}}}\psi , 
\end{equation}
where both $V_{\rm{SN}}$ and $\gamma_{\rm SN}$ are model parameters. The circular velocity used is that at the half-mass radius of the disk for quiescent star formation or the circular velocity at the half-mass radius of the spheroid for starbursts. Separate values of $V\textsubscript{SN}$ can be specified for star formation taking place in the disk ($V\textsubscript{SN, disk}$) and bulge ($V\textsubscript{SN, burst}$) components of a galaxy.
These parameters have generally been assumed to be the same in most previous versions of the model (although see \citealt{Benson2010b} for a counter-example). Gas ejected from the halo is assumed to return from a reservoir beyond the halo's virial radius to the hot gas reservoir in the main halo at a rate given by
\begin{equation}
    \dot{M}\textsubscript{return} = \alpha\textsubscript{ret}\frac{M\textsubscript{res}}{\tau\textsubscript{dyn,halo}} ,
\end{equation}
where $\tau\textsubscript{dyn,halo}$ is the dynamical time of the halo, $M\textsubscript{res}$ is the mass in the reservoir of ejected gas beyond the virial radius, and $\alpha\textsubscript{ret}$ is a free parameter.

\subsection{SMBH growth and AGN feedback}

Supermassive black holes (SMBH) can inject energy into the halo gas, inhibiting gas cooling. Hot halo accretion, BH-BH mergers, as well as starbursts, can all increase the mass of the black hole \citep{Bower2006, Croton2006, Lagos2008}. In the case of starbursts, the mass accreted onto the SMBH is a fraction $f\textsubscript{SMBH}$ of the mass of stars formed, where $f\textsubscript{SMBH}$ is an adjustable parameter. AGN heating is assumed to occur if both of the following conditions are met: (1) the gas halo is in quasi-hydrostatic equilibrium, that is the condition:
\begin{equation}
    \tau\textsubscript{cool}/\tau\textsubscript{ff} > 1/\alpha\textsubscript{cool} ,
\end{equation}
is met, where $\tau\textsubscript{cool}$ is the cooling time of the gas, $\tau\textsubscript{ff}$ is the free-fall time, and $\alpha\textsubscript{cool}$ is an adjustable parameter; (2) the AGN power required to balance the radiative cooling luminosity of the hot halo gas is below a fraction $f\textsubscript{Edd}$ of the Eddington luminosity of the SMBH (which is held fixed at $f\textsubscript{Edd}=0.01$). The implementation of the growth of SMBH in \texttt{GALFORM} is described in \cite{Lacey2016} and \cite{Griffin2019}.

\subsection{Stellar initial mass function}

The stellar initial mass function (IMF) gives the mass distribution of newly formed stars and strongly affects the evolution of the total luminosity, as well as the metal and gas content of the galaxy. The IMF, $\Phi(m)$ is defined such that the number of stars with mass $m$ is $dN = \Phi(m)\text{dln}m$, and $\Phi(m)$ is normalised such that $\int_{m_{\rm L}}^{m_{\rm H}} m \Phi(m) \text{dln}m = 1$ between some minimum ($m_{\rm L}$) and maximum ($m_{\rm H}$)  stellar mass.

In some of the models considered later, we will allow the slope of the IMF to depend on the mode of star formation, assuming a solar neighbourhood IMF for quiescent star formation that takes place in disks and a power law IMF for bursts of star formation. The slope of the IMF power law, $x$,  is then a model parameter:  
\begin{equation}
    \Phi(m) := dN/d\text{ln}m \propto m^{-x}.
\end{equation}
To accommodate this change, we must self-consistently calculate the recycled fraction (i.e. the fraction of mass returned to the ISM from mass lost by stars over their lifetime), given by 
\begin{equation}
    R = \int_{1M{\odot}}^{m_{H}}(m-m\textsubscript{rem}(m))\Phi(m) {\rm d}\text{ln}m,
\end{equation}
where $m\textsubscript{rem}$ is the mass of the remnant left by a dying star of birth mass $m$. We also calculate the yield, $p$, i.e. the fraction of the initial mass synthesised into metals and ejected,  as
\begin{equation}
    p = \int_{1M{\odot}}^{m_{H}}p\textsubscript{Z}(m) m \Phi(m) {\rm d }\text{ln}m,
\end{equation}
where $p\textsubscript{Z}(m)$ is the fraction of mass ejected as newly produced metals by a star of mass $m$.

For reference, the solar neighbourhood IMF  \citep{Chabrier2003} assumed in quiescent star formation tends to a power law slope of $x=1.35$ above one solar mass and turns over with a log-normal form below one solar mass. 
Most published variants of \texttt{GALFORM} have adopted a solar neighbourhood IMF in all modes of star formation, typically the Kennicutt IMF \citep{Kennicutt:1983}. Different forms of the IMF are compared in fig.~2 of \cite{Lacey2016}.

\subsection{Stellar Population Synthesis}

We calculate galaxy spectral energy distributions (SEDs) by building a composite stellar population from simple stellar populations (SSPs) of age $t$ and metallicity $Z$, for the adopted IMF,  using the predicted star formation and chemical evolution histories of the progenitors of a galaxy. To calculate the SED for each SSP we use the stellar population synthesis (SPS) model developed by \cite{Conroy2009}. 

We consider two variants of \texttt{GALFORM}. In the first case, we assume that the IMF is the same in starbursts and quiescent star formation, and has the solar neighbourhood form proposed by \cite{Chabrier2003}. In the second case, quiescent star formation again takes place following a solar neighbourhood IMF, but the IMF in bursts takes on a power-law form, with the slope of the power law being a model parameter. The integral to find the luminosity per unit wavelength $L_{\lambda}$ is performed separately over stars formed in bursts and those formed in the disk. The results for the disk and bulge are then added to get the total galaxy luminosity.

The \texttt{FSPS} model  \citep{Conroy2013} includes a flexible treatment for the contribution to the SED from thermally-pulsating asymptotic giant branch (TP-AGB) stars, a stellar phase which is difficult to model accurately. \cite{Conroy2013} introduces a parameterisation to adjust the effect of the TP-AGB contribution. Adjusting these parameters, we find that this has little effect on the datasets that we consider, so we use the default values. \cite{Gonzalez-Perez2014} show the effect of choosing different SPS models on the \texttt{GALFORM} model predictions. Gonzalez-Perez et~al. found that the model predictions for the luminosity function in the ultra-violet and optical are insensitive to the choice of the near-infrared SPS model, whereas the evolution of the luminosity function in the near-infrared does depend on the treatment of TP-AGB stars. The \texttt{FSPS} model allows new SSP tables to be generated for any IMF.

\subsection{Absorption and reradiation of starlight by dust}

Within galaxy disks, a two-component dust model is assumed with diffuse and molecular cloud components. In both cases, the dust is mixed in with the stars. 
The mass of dust in a galaxy is assumed to be a constant fraction $\delta\textsubscript{dust}$ of the mass of metals in the cold gas, $M\textsubscript{dust} = \delta\textsubscript{dust}Z\textsubscript{cold}M\textsubscript{cold}$, where $\delta\textsubscript{dust} = 0.334$ \citep{Silva1998}. The dust grain size distribution is assumed to be the same as in the solar neighbourhood. The optical depth of dust in the disk is then calculated as 
\begin{equation}
    \tau_{\rm{dust}, \lambda} =  0.043\left(\frac{k_{\lambda}}{k\textsubscript{V}}\right)    
    \left(\frac{\Sigma\textsubscript{gas}}{M_{\odot}\text{pc}^{-2}}\right)    
    \left(\frac{Z\textsubscript{cold}}{0.02}\right),
\end{equation}
where $\Sigma\textsubscript{gas}$ is the surface density of the gas, ${k_{\lambda}}$ represents the extinction curve (being the wavelength-dependent extinction coefficient normalized to some reference wavelength), and $k\textsubscript{V}$ is the extinction coefficient in the V-band, which is used as the normalization reference for ${k_{\lambda}}$. 

The model assumes that a fraction $f\textsubscript{cloud}\,$ of the gas and dust is in clouds of mass $m\textsubscript{cloud} = 10^{6} M_{\odot}$ and radius $r\textsubscript{cloud} = 16 \,\textrm{pc}$, based on observations of local galaxies \citep{Granato2000}. Stars are assumed to form inside molecular clouds and escape over an adjustable timescale $t\textsubscript{esc}$. The optical depth in each cloud therefore scales as $\tau\textsubscript{cloud} \propto m\textsubscript{cloud}/r^{2}\textsubscript{cloud}$, and the optical depth of the diffuse component as $\tau\textsubscript{diffuse}\propto(1-f\textsubscript{cloud})M\textsubscript{cold}Z\textsubscript{cold}/r^{2}\textsubscript{diffuse}$. Here, $r\textsubscript{diffuse}$ is taken to be the disk radius, $r\textsubscript{disk}$, for stars in the disk, and $r\textsubscript{bulge}$, for stars in the bulge. Attenuation by diffuse dust is calculated by interpolating the tabulated results of radiative transfer runs by \cite{Benson2018} using the \texttt{HYPERION} code of \cite{Hyperion}. These models are higher resolution versions of the dust extinction tables previously used in \texttt{GALFORM}, which were based on the radiative transfer calculations by \cite{Ferrara1999}, and extend over a wider range of optical depth values. 

The dust re-radiates the energy it absorbs from the starlight at infrared and sub-millimetre wavelengths. In the model, the emission from the diffuse and cloud components are treated separately. We calculate the total stellar luminosity absorbed by the dust in a galaxy, and assume the dust reradiates this energy as a modified black body:
\begin{equation}
    L_{\lambda}^{\text{dust}} \propto M\textsubscript{dust}\kappa\textsubscript{d}(\lambda)B_{\lambda}(T\textsubscript{dust}),
\end{equation}
where $M\textsubscript{dust}$ is the mass and $T\textsubscript{dust}$ the temperature of the dust component, $B_{\lambda}(T)$ is the Planck function, and $\kappa\textsubscript{d}$ is the dust opacity per unit mass. Integrating $L_{\lambda}^{\text{dust}}$ over wavelength and setting the result equal to the absorbed luminosity allows us to solve for the dust temperature, $T\textsubscript{dust}$, for each component. The dust opacity in the IR/sub-mm is approximated as a broken power law 
\begin{equation}
\kappa\textsubscript{d} \propto \begin{cases} 
      \lambda^{-2} & \lambda < \lambda\textsubscript{b} \\
      \lambda^{-\beta_{\text{b}}} & \lambda > \lambda\textsubscript{b},  
   \end{cases}
\end{equation}
where we allow the adjustable exponent $\beta\textsubscript{b}$ for bursts to vary within the range 1.5 - 2.0 (see e.g. \citealt{Silva1998}). For quiescent disks, $\lambda\textsubscript{b} = \infty$ and for bursts $\lambda\textsubscript{b} = 100 \, \mu \textrm{m}$.

\section{Bayesian optimisation}

We use Bayesian optimisation (see \citealt{Frazier2018} and \citealt{garnett_bayesoptbook_2023} for reviews) to set the model parameters so that \texttt{GALFORM} reproduces as closely as possible the observational datasets used in the model calibration. Here we explain why we take this approach and set out its background. We start in \S~3.1 with an overview of the problem, the exploration of a high dimensional parameter space of a computationally expensive model, and explain how Bayesian optimisation helps us to address this challenge. The nature and role of Gaussian processes, the tool used to encapsulate our knowledge of the parameter space is described in \S~3.2. The kernel function is an important part of the Gaussian process description and is explained in \S~3.3. The method used to choose where in the parameter space to add new calculations with the full model is described in \S~3.4. The datasets used to calibrate \texttt{GALFORM} are listed in \S~3.5. A simple validation of our approach is given in \S~3.6. We close in \S~3.7 by describing the practical application of Bayesian optimisation to \texttt{GALFORM}.

\subsection{An overview of Bayesian optimization}
\label{sec:bo_overview}

Calibrating a galaxy formation model involves finding parameter values that make the model predictions match observations as closely as possible. We measure this match using a metric, $f$, which quantifies the discrepancy between model predictions and observational data---the smaller the value of $f$, the better the match.

In a traditional calibration approach, we would evaluate the model at many different points in the parameter space (where each point represents a specific choice of parameter values) and rank them according to $f$. For a small number of parameters, we could simply try every possible combination on a grid. However, in the application here \texttt{GALFORM} has 15 adjustable parameters, making a complete grid search computationally impossible. Even sophisticated methods like Markov Chain Monte Carlo (MCMC; \citealt{MCMC}) require thousands of model evaluations to thoroughly explore such a high-dimensional space.

The fundamental challenge is that each \texttt{GALFORM} evaluation has a modest computational cost that soon becomes prohibitive if a large number of runs is needed. For the predictions used here, we must run the model at many different redshifts to obtain galaxy number counts and redshift distributions, making each evaluation costly. We need a method that can find good parameter values with far fewer model runs.

Bayesian optimisation provides a solution by using a scheme to decide where to evaluate the model next. Instead of blindly sampling the parameter space, it builds up knowledge about the metric $f$ across the entire parameter space and uses this knowledge to decide where the next evaluation is most likely to improve our understanding or find a better fit. A simple analogy is looking for a missing set of keys: think of it as the difference between randomly searching for your keys versus systematically checking the places they are most likely to be, and updating your search strategy based on where you have already looked.

The key insight is that we do not need to know the exact value of $f$ everywhere in the parameter space. Instead, we maintain a probabilistic description of what $f$ might be at each location, with varying degrees of certainty. Where we have evaluated the full model, our certainty is high. Where we have not explored, our uncertainty is larger. This probabilistic description is provided by a Gaussian process (GP), which we explain in the next subsection.

After an initial set of model evaluations (typically about 30 for our 15-dimensional space, obtained using Latin hypercube sampling), Bayesian optimisation begins an iterative process: (1) use the GP to estimate $f$ everywhere in the parameter space, (2) identify where the next evaluation would be most valuable, (3) run \texttt{GALFORM} at that location, and (4) update the GP with this new information. As we show later, this approach typically finds excellent parameter values within just 100--200 model evaluations---orders of magnitude fewer than traditional methods would require.

This is technically classified as a black-box, derivative-free, global optimisation problem \citep{Frazier2018}. ``Black-box'' means we do not assume any particular mathematical form for $f$; ``derivative-free'' means we do not compute how $f$ changes with respect to each parameter (information that could guide a gradient-based search); and ``global optimisation'' means we aim to find the best solution across the entire parameter space, not just a local minimum.

\begin{figure*}
    \centering
    \includegraphics[width=0.99\textwidth]{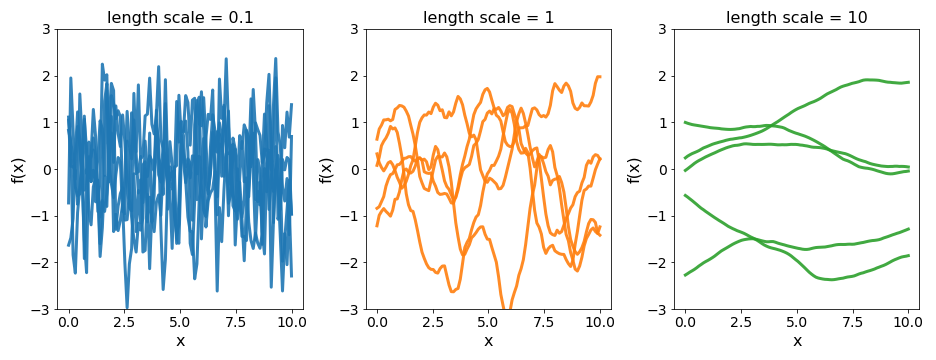}
    \vspace{-0.5cm}
    \caption{A demonstration of the effect of the length scale adopted in the kernel function on the appearance of a Gaussian process (GP). Each panel shows several realisations or draws from a GP. In each case, the process has zero mean. However, the hyperparameter that governs the scales over which values of $f$ are correlated varies between panels. The left panel shows the shortest correlation length scale, with $\theta=0.1$, the middle panel shows 1.0, and the right panel 10.0. A shorter length scale corresponds to a function which changes rapidly with small changes to the input parameters.}
    \label{fig:length_scale_example}
\end{figure*}

\begin{figure*}
    \centering
    \includegraphics[width=0.97\textwidth]{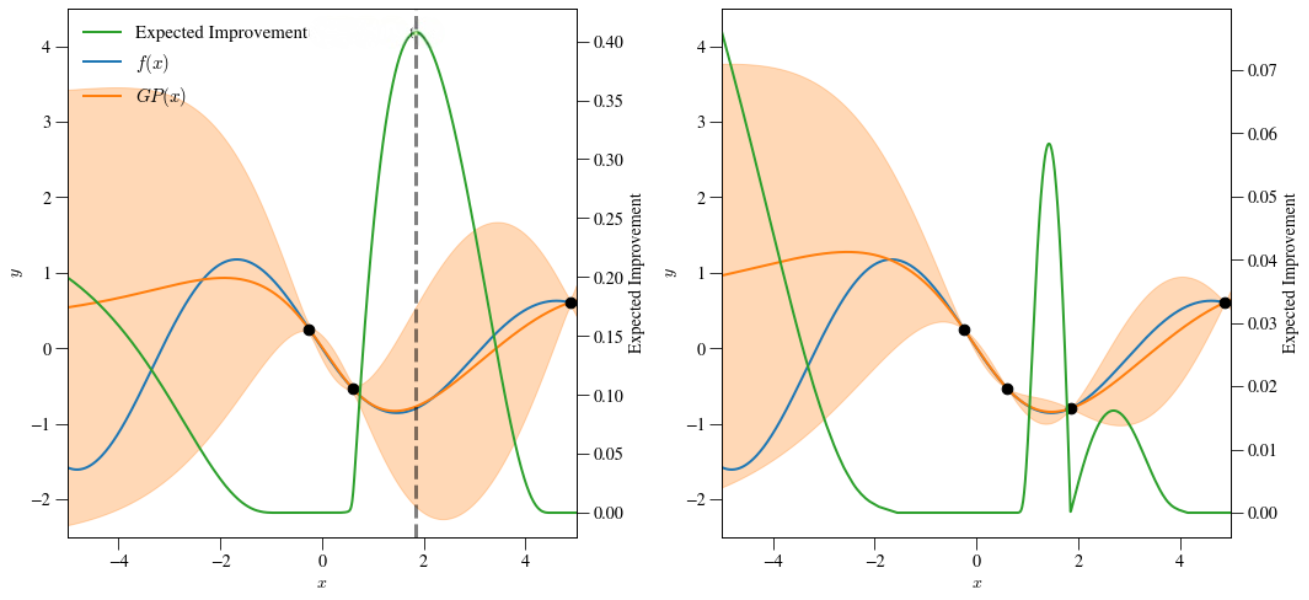}
    \caption{An illustration of one iteration of the expected improvement (EI) algorithm. The objective is to find the minimum value of the target function shown by the blue curve. Two iterations are plotted, $n=3$ (left panel) and $n=4$ (right panel). 
    The Gaussian process (GP) posterior (orange solid line) is an estimate of the target function. The orange-shaded region shows the $3\sigma$ confidence interval of the GP. The left panel shows the GP after $3$ evaluations of the function (shown by the black solid points).  The green curve shows the EI (right axis), which corresponds to the expectation integral of the GP posterior below the minimum evaluation so far (i.e. how much we expect to improve upon the current minimum evaluation at each point $x$). The EI suggests that a new evaluation just below $x=2$ will bring the biggest improvement in our knowledge of the target function. The right panel shows the updated GP posterior and EI curve after evaluating the function for $n=4$ at the point of maximum expected improvement, as shown by the black dashed line in the left panel. At this point, the next evaluation would be chosen to be below $x=-4$.}
    \label{fig:gaussian_process_example}
\end{figure*}

\subsection{Gaussian processes: describing our knowledge of the parameter space}
\label{sec:gp}

A Gaussian process (GP) provides a mathematical framework for describing what we know about the metric function $f$ throughout the parameter space \citep{GP}. Crucially, the GP not only gives an estimate of $f$ at any location, but also quantifies our uncertainty in that estimate.

To understand GPs intuitively, imagine you are trying to map the elevation of a landscape. If you have measured the height at a few specific locations, you can make reasonable guesses about the height at nearby points---elevation typically changes smoothly rather than jumping erratically. Points very close to your measurements can be estimated with high confidence, while points far from any measurement are more uncertain.

Quantitatively, we consider the function $f(\mathbf{x})$, where $\mathbf{x}$ represents a specific set of values for all 15 parameters. The GP assumes that the joint probability distribution for the values of $f$ at a set of $n$ points in this parameter space is a multivariate Gaussian, with mean $\mu(\mathbf{x})$, and covariance $K(\theta,\mathbf{x_i},\mathbf{x_j})$ between points $\mathbf{x_i}$ and $\mathbf{x_j}$ that also depends on parameters $\theta$. Mathematically, this can be written as $f \sim \mathcal{N}(\mu,K)$. In our application, we set the unconstrained mean $\mu(\mathbf{x})$ to zero because our target is to minimize $f$ (make it as close to zero as possible). If the values of $f$ at a set of points $\mathbf{x_1},\ldots \mathbf{x_n}$ are already known (through exact evaluations of the model), then the probability distribution for the value of $f$ at a point $\mathbf{x}'$ is given by the conditional probability distribution. This conditional probability distribution is also a Gaussian, with constrained mean $\mu(\mathbf{x}')$ and constrained variance $\sigma^2(\mathbf{x}')$ that depend on the values of $f$ at the points $\mathbf{x_1},\ldots \mathbf{x_n}$. The mean value $\mu(\mathbf{x'})$ is our best estimate of $f(\mathbf{x'})$, while $\sigma(\mathbf{x}')$ measures the uncertainty in this estimate.

The covariance matrix $K(\theta, \mathbf{x_i}, \mathbf{x_j})$ encodes the key assumption that underlies GPs: nearby points in the parameter space should have similar values of $f$. This matrix describes how strongly the value of $f$ at one location $\mathbf{x_i}$ is correlated with its value at another location $\mathbf{x_j}$. The closer two points are in the parameter space, the more strongly correlated their $f$ values should be. The parameter $\theta$ controls the length scale over which correlations persist---essentially, how rapidly $f$ can change as we move through the parameter space.

To illustrate how the length scale affects the GP, Fig.~\ref{fig:length_scale_example} shows several random draws from GPs with different correlation lengths. Each curve represents a possible realization of what $f$ might look like. When the correlation length is short (left panel), the function varies rapidly. When it is long (right panel), the function varies smoothly over larger distances. The GP learns appropriate length scales from the data.

As we evaluate \texttt{GALFORM} at more points in the parameter space, the GP is updated with this new information. The uncertainty shrinks to nearly zero at the locations where we have run the full model, while remaining larger in unexplored regions. This updated GP then guides where we should evaluate the model next, as we describe in Section~\ref{sec:acquisition}.

\subsection{The kernel function: encoding smoothness}
\label{sec:kernel}

The kernel function (also called the covariance function) is the mathematical tool that implements the correlation structure of the GP. It determines how the value of $f$ at one location in the parameter space relates to its value at another location.

For our application, we use the Mat\'ern 5/2 kernel, which has proven effective for optimisation problems in many fields. The covariance between two points $\mathbf{x}_i$ and $\mathbf{x}_j$ in the parameter space is given by:
\begin{equation}
K(\mathbf{x}_i, \mathbf{x}_j) = \sigma^2 \left(1 + \sqrt{5}r + \frac{5r^2}{3}\right) \exp\left(-\sqrt{5}r\right),
\label{eq:matern}
\end{equation}
where
\begin{equation}
r = \sqrt{\sum_{m=1}^{d} \frac{(x_{im} - x_{jm})^2}{l_m^2}}.
\label{eq:distance}
\end{equation}

Here, $d$ is the number of dimensions in the parameter space (15 in our case), $x_{im}$ denotes the value of parameter $m$ at location $\mathbf{x}_i$, and $l_m$ is the length scale for parameter $m$. The parameter $\sigma^2$ controls the overall variance.

A crucial feature of this kernel is that it includes a separate length scale $l_m$ for each parameter dimension. This property, called \emph{automatic relevance determination}, allows the GP to learn that changes in some parameters affect $f$ more strongly than changes in others. Parameters with long length scales contribute less to determining $f$---the metric changes slowly as these parameters vary. Conversely, parameters with short length scales are highly relevant---small changes in these parameters lead to significant changes in how well the model matches observations.

The Mat\'ern 5/2 kernel has several advantages for our application:
\begin{itemize}
    \item It assumes that $f$ is smooth (twice differentiable), which is physically reasonable for our model
    \item It is flexible enough to capture complex behaviour
    \item It is computationally efficient to evaluate
    \item The automatic relevance determination helps identify which parameters most strongly affect the model fits
\end{itemize}

By learning appropriate length scales from the \texttt{GALFORM} evaluations, the GP builds an increasingly accurate representation of the metric function across the parameter space.

\subsection{Choosing where to evaluate next: the acquisition function}
\label{sec:acquisition}

At each step of the optimisation, we must decide where in the parameter space to evaluate \texttt{GALFORM} next. This decision involves a fundamental trade-off. On one hand, we want to evaluate the model where we expect to find parameter values that give a better fit to the data (lower $f$). On the other hand, we want to explore regions where our uncertainty is high, as these unexplored regions might contain even better solutions that we have not yet discovered. This is the classic \emph{exploitation versus exploration} dilemma.

The acquisition function provides a mathematical framework for balancing this trade-off. It assigns a score to every possible location in the parameter space, indicating how valuable it would be to evaluate the model there. We then choose to evaluate \texttt{GALFORM} at the location with the highest acquisition function value.

We use the \emph{expected improvement} (EI) acquisition function, which has a clear intuitive interpretation. Let $f_{\rm min}^n$ denote the smallest (best) value of the metric we have found after $n$ evaluations of \texttt{GALFORM}. For any new location $\mathbf{x}$, we can ask: what improvement over $f_{\rm min}^n$ can we expect if we evaluate the model at $\mathbf{x}$? The EI quantifies this expected improvement, accounting for both the GP's predicted value of $f(\mathbf{x})$ and the uncertainty in this prediction.

More formally, we define a utility function, $u$:
\begin{equation}
u(\mathbf{x}) = \max\left(0, f_{\rm min}^n - f(\mathbf{x})\right).
\label{eq:utility}
\end{equation}
This utility is zero if the new evaluation would not improve upon our current best result (i.e. $f(\mathbf{x}) >f_{\mathrm{min}}^{n}$), and positive if it would yield a better fit. Since the GP provides a probability distribution for $f(\mathbf{x})$ at any location (rather than a single value), we compute the expected utility by integrating over all possible values weighted by their probabilities:
\begin{equation}
a_{\rm EI}(\mathbf{x}) = \mathbb{E}[u(\mathbf{x})] = \int_{-\infty}^{\infty} u(\mathbf{x})\,{\rm GP}(\mathbf{x})\,{\rm d}\mathbf{x},
\label{eq:ei}
\end{equation}
where ${\rm GP}(\mathbf{x})$ denotes the probability distribution given by the Gaussian process at location $\mathbf{x}$.

Fortunately, this integral has a closed-form solution. Let $\mu(\mathbf{x})$ and $\sigma(\mathbf{x})$ denote the conditional mean and conditional standard deviation predicted by the GP at location $\mathbf{x}$. Define the standardized improvement:
\begin{equation}
Z(\mathbf{x}) = \frac{f_{\rm min}^n - \mu(\mathbf{x})}{\sigma(\mathbf{x})},
\label{eq:z}
\end{equation}
assuming $\sigma(\mathbf{x}) > 0$. Then the expected improvement can be written as:
\begin{equation}
a_{\rm EI}(\mathbf{x}) = \sigma(\mathbf{x})\left[Z(\mathbf{x})\Phi(Z(\mathbf{x})) + \phi(Z(\mathbf{x}))\right],
\label{eq:ei_closed}
\end{equation}
where $\Phi$ is the cumulative distribution function of the standard normal distribution and $\phi$ is its probability density function.

This formula (Eqn. 20) explicitly reveals how exploitation and exploration contribute to the expected improvement. Both terms inside the brackets are multiplied by the uncertainty $\sigma(\mathbf{x})$, which means that \emph{both contributions vanish in regions where we are certain about $f$}:
\begin{itemize}
    \item \textbf{Exploitation contribution}: The term $Z(\mathbf{x})\Phi(Z(\mathbf{x}))$ is large when the GP predicts a mean value $\mu(\mathbf{x})$ much lower than the current best $f_{\rm min}^n$ (making $Z$ large and positive). The cumulative distribution function $\Phi(Z)$ gives the probability that sampling at $\mathbf{x}$ will yield an improvement. This term favors regions where we expect good results.
    \item \textbf{Exploration contribution}: The term $\phi(Z(\mathbf{x}))$ is the standard normal probability density, which is largest when $Z \approx 0$---that is, when the predicted mean $\mu(\mathbf{x})$ is close to the current best value. This term ensures that regions with high uncertainty are explored even when the predicted mean is only moderately promising.
\end{itemize}

Crucially, the factor $\sigma(\mathbf{x})$ outside the brackets means that \emph{both} exploitation and exploration are scaled by uncertainty. At points we have already evaluated, where $\sigma(\mathbf{x}) = 0$, the EI is exactly zero---there is no value in re-evaluating the model at the same parameters, regardless of how good those parameters were. Conversely, regions with large uncertainty receive higher scores, even if the mean prediction alone would not make them the most promising location.

The optimal balance between exploitation and exploration emerges automatically from this mathematical structure. A region with very low $\mu(\mathbf{x})$ (good predicted performance) but small $\sigma(\mathbf{x})$ (we are already confident about this region) may score lower than a region with moderate $\mu(\mathbf{x})$ but large $\sigma(\mathbf{x})$ (significant chance of discovering something unexpectedly better).

Fig.~\ref{fig:gaussian_process_example} illustrates how the EI acquisition function works for a simple one-dimensional example. In the left panel, we show an unknown target function (blue curve) that we are trying to minimize. We have evaluated this function at three locations (black points). The orange line shows the GP's current estimate of the function, with the shaded region indicating the uncertainty (specifically, the $3\sigma$ confidence interval). Notice that the uncertainty is zero at the evaluated points and grows larger in unexplored regions.

The green curve shows the EI acquisition function. It is highest slightly to the right of $x=2$, balancing two factors: (1) the GP predicts relatively low values of $f$ in this region (exploitation), and (2) the uncertainty is large enough that we might discover even lower values (exploration). The black dashed line marks where we would evaluate the function next.

The right panel shows the updated GP and EI after evaluating the function at the location identified in the left panel. The uncertainty has decreased around the new evaluation point, and the EI now identifies a different location (far left) as the most promising for the next evaluation.

This is an iterative process: \texttt{GALFORM} is evaluated at the location with the highest EI, the GP is updated with the new result, the EI is recalculated across the parameter space, and the next location at which to evaluate the model is found. After sufficient iterations, the algorithm converges to a parameter set that gives the best fit to the observational data, and which subsequently is not significantly improved upon, as we demonstrate in Section~\ref{sec:validation}.

\subsection{Dataset selection for parameter calibration}
\label{Dataset selection}

We aim to test the ability of  \texttt{GALFORM} to reproduce observations of dusty star-forming galaxies, SMGs, detected at sub-mm wavelengths, which tend to be high-redshift objects, as well as matching the properties of the low-redshift galaxy population. In \cite{Elliott} we used nine observational datasets measured for local galaxies to calibrate \texttt{GALFORM}. To make the analysis simpler to follow, here we restrict ourselves to the $z=0$ $K$-band luminosity function and supplement this with the number counts and redshift distribution of SMGs at $850\, \mu$m. We investigate if the best fitting models can reproduce the local $850\mu\text{m}$ luminosity function and other selected $z=0$ galaxy observations in \S~4.3, but emphasize that these datasets are not used in the calibration. 

Below we list the observational datasets used to constrain the model parameters: 
\begin{itemize}
    \item For the local $K$-band LF, we compare to the observational estimate from \cite{Kochanek2001}.
    \item For the SMG number counts, at the bright end we compare to \cite{Stach2018}, based on ALMA $870\ \mu$m follow-up observations of a sample of sources detected at $850\ \mu$m by SCUBA,
    and at the faint end we compare to \cite{Chen2013}. 
    \item For the SMG redshift distribution, we compare the model to the measurements from \cite{Ugne2020}. These authors used the same $850\ \mu$m SCUBA and $870\ \mu$m ALMA data as \citet{Stach2018}, and then used ancillary data at other wavelengths in conjunction  with the MAGPHYS code \citep{DaCunha2015} to obtain photometric redshifts. We apply a flux limit $S\textsubscript{870}>4.0$~mJy (as in \citealt{Stach2018}) to the \cite{Ugne2020} sample, and then weight the individual sources by the scaling factors as a function of $850\ \mu \text{m}$ flux from \cite{Geach2017}, which account for varying depth and incompleteness in the original SCUBA survey. We then calculate the redshift distribution by stacking the MAGPHYS-derived redshift probability distributions for the individual sources. 

\end{itemize}

\begin{figure}
    \includegraphics[width=\columnwidth,trim={2.3cm 17.3cm 4cm 2.8cm}, clip]{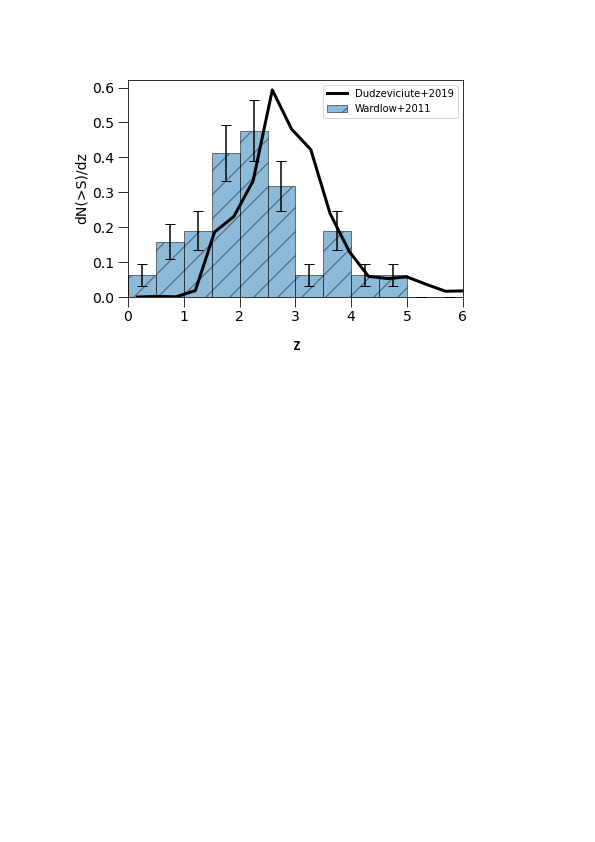}
    \caption{Comparison between the redshift distribution for SMGs brighter than 4 mJy inferred by \protect\cite{Ugne2020}  (black solid line) and \protect\cite{Wardlow2011} (blue histogram, which  only includes SMGs with robust optical counterparts). Both redshift distributions have been normalized to unit area under the curve. Here, we calibrate \texttt{GALFORM} to the redshift distribution estimated by \protect\cite{Ugne2020}.} 
    \label{fig:smg_dist_comparison}
\end{figure}

It is important to note that the redshift distribution of SMGs that we obtained using the data from  \cite{Ugne2020} has a somewhat \textit{higher} median redshift than previous estimates used to calibrate the \texttt{GALFORM} models. 
\cite{Wardlow2011} reported a median photometric redshift for their sample of $z = 2.2 \pm 0.2$ when including only SMGs with robustly detected optical counterparts, 
but this increased to $z=2.5 \pm 0.5$ when a slightly higher flux limit  $S\textsubscript{850}>4$mJy was applied and allowing for the likely redshifts of SMGs with no identifiable optical counterpart.
For comparison, applying the same flux limit $S\textsubscript{850}>4$mJy to the sample of \cite{Ugne2020} gives a median redshift of $z = 2.8 \pm 0.09$
The two estimates of the SMG redshift distribution for sources brighter than 4 mJy are shown in Fig.~\ref{fig:smg_dist_comparison} (where the distribution plotted for the \cite{Wardlow2011} sample only includes SMGs with robust optical identifications). The model calibrations performed in \cite{Lacey2016} and \cite{Baugh2019} thus used an SMG redshift distribution with a somewhat lower median redshift than the calibration data employed here (note that Lacey et~al., applied a brighter flux limit to the SMG redshift distribution, but only considered SMGs from Wardlow et~al. with optical counterparts). It is therefore interesting to find out if the model with a top-heavy IMF can still reproduce the higher median redshift found by \cite{Ugne2020}, and if this has implications for the value of the recovered slope of the IMF in starbursts. 

We will see later that the \texttt{GALFORM} predictions for the redshift distribution of SMGs show a series of spikes. The spikes at intermediate and high redshifts arise due to the limited number of halo merger trees used in the calculation. Using many more halos would smooth out these spikes, but at the expense of greatly increasing the computational cost for each full model evaluation. The spike at low redshift ($z \sim 0.1$) is different, being robust to using more halos, and results from the local population of quiescently star-forming galaxies. This local SMG population appears to be under-represented in the SMG sample analysed by \cite{Ugne2020}, possibly because the survey field was chosen to avoid nearby galaxies so as to target the high-redshift population. However, the local SMG population is robustly detected observationally in surveys that measure the local $850\ \mu$m LF, as discussed in \S\ref{sec:further_predictions}.
To remove any issue with the low redshift spike, we excluded both both model and observed SMG at $z<0.8$ when we calculated the MAE for the SMG redshift distribution in the parameter fitting.

\subsection{Validation of the optimisation approach}
\label{sec:validation}

Before running the Bayesian optimisation on the  \texttt{GALFORM} model, we first assess whether this method is likely to  succeed. We also need to gain some insight into  the convergence properties of the optimisation process (or parameter calibration) so that we have some guidance as to when to end the search. To do this, we build an approximate emulator of \texttt{GALFORM} so that we can get a better understanding of the optimisation routine, without requiring many time-consuming evaluations of the \texttt{GALFORM} model. 

To carry out this test, we first build a neural network emulator to predict all three observational datasets described in \S\ref{Dataset selection} from 500 runs of the full \texttt{GALFORM} model, with samples drawn from the space defined by the parameter ranges given in Table~\ref{tab:ParamRanges} using a Latin hypercube. The trained neural network provides us with a fast alternative to running \texttt{GALFORM}, and allows us to test our optimisation strategy before applying it to the full model. This is similar to the approach taken in \cite{Elliott}. 
One might ask why we cannot simply use this emulator to perform the model calibration instead of using Bayesian optimisation. The observables considered here are more complicated for the neural network to learn than those used to calibrate the model in Elliott et al.; in particular, the redshift distribution of SMGs is hard to reproduce accurately, without running many more full \texttt{GALFORM} evaluations. Hence, if we used the trained neural network to give a set of parameters deemed to be the best fitting ones, these may not lead to the full \texttt{GALFORM} model reproducing the calibration data closely. Also, as commented above, because we are predicting statistics that are computed from the model output at many redshifts, each model evaluation has a higher computational overhead than in Elliott et al. Hence, whilst the emulator is too approximate to use to calibrate \texttt{GALFORM}, it has similar properties to the model and serves to let us test the optimisation process. 

\begin{figure}
    \centering
    \includegraphics[width=0.98\columnwidth]{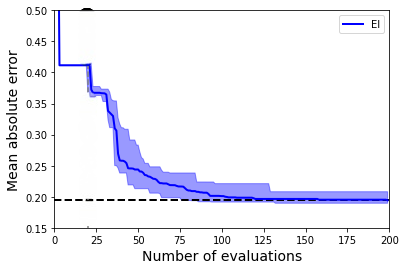}
    \caption{Performance of the Expected Improvement (EI) Bayesian Optimisation algorithm on a neural network emulator of \texttt{GALFORM}. The solid blue line shows the median over 30 separate runs. The shaded region shows the minimum to maximum range for these runs. The dashed horizontal line shows the global minimum found using MCMC, with 20 chains of 10 000 steps each.}
    \label{fig:example}
\end{figure}

Here, we use a neural network model from the \texttt{TensorFlow} software library \citep{tensorflow}. From our full \texttt{GALFORM} runs, we have sets of parameters $\textbf{x}$, and an associated output $\textbf{y}$, which corresponds to the \texttt{GALFORM} predictions for the datasets we are considering (for example, the values of the $K$-band luminosity function in different magnitude bins). We then calculate the error or distance between the \texttt{GALFORM} predictions and the calibration datasets as given by the mean absolute error, MAE, which for a single dataset is 
\begin{equation}
    \text{MAE} = \frac{1}{n}\sum_i  | y_{i} - y_{\text{obs}, i} |,
\end{equation}
where $y\textsubscript{obs}$ corresponds to the observational calibration dataset which has $n$ bins. Importantly, we linearly rescale each dataset so that $y_{\textrm{obs}}$ lies within the range [0,1], so that we can combine the errors defined in this way for different datasets in a consistent way into a single distance value or metric. In the case of the $K$-band LF and the SMG number counts it is the logarithm of these quantities that is rescaled. For the SMG redshift distribution, we normalise both predicted and observed distributions over the same redshift range (here taken to be $z>0.8$).
We then calculate the total MAE to be minimised as a weighted mean of the MAEs for the individual observational datasets. This allows us to give more weight to some datasets than others, if desired.

We train the neural network to predict the error between the \texttt{GALFORM} predictions and the observational data, given the parameter values $\textbf{x}$ so that we have a fast-to-evaluate but approximate emulator of \texttt{GALFORM} on which to test our Bayesian optimisation strategies. By building an emulator which we can evaluate quickly, we allow ourselves to find the approximate global minimum error in a short amount of time using MCMC.  Of course, the emulated model will not correspond exactly to running the full \texttt{GALFORM} model, but it will be of similar complexity and so allows us to assess the optimisation methods, ready to apply to the full \texttt{GALFORM} model.

Having trained our neural network emulator to produce an approximate error between the \texttt{GALFORM} predictions for the $K$-band LF, the SMG redshift distribution and number counts, and the corresponding observational datasets, given a set of input parameters, we first find an approximate global optimum using an extensive MCMC search. We use a simple implementation of the Metropolis-Hastings algorithm  (e.g. \citealt{MCMC}), using 20 chains, each 10\,000 steps in length and starting from a randomly chosen location in the parameter space, to locate the set of parameters which returns the lowest approximate error for the neural network emulator, as judged by the MAE metric. 

Next, we test our optimisation strategies to see if we can find a set of best-fitting parameters that gives a comparable error or MAE to that obtained with MCMC. To do this, we initially perform a pseudo-random sampling of the parameter space using Sobol sampling (e.g. see \citealt{Oleskiewicz2020}), drawing 2 samples per dimension of the parameter space to be searched (in our case 30 initial evaluations for a 15 dimension parameter space). We evaluate the neural network emulator at these points to return approximate errors, fit the GP to the parameter-error pairs, and begin the expected improvement search to decide where to evaluate the neural network emulator next in the parameter space.

Fig.~\ref{fig:example} shows the result of applying the EI Bayesian optimisation to the neural network emulator of \texttt{GALFORM}. The $y$-axis gives the MAE returned by the neural network, with the $x$-axis showing the total number of evaluations of the emulator carried out up to that point. The first 30 runs are chosen through pseudo-random Sobol sampling. The Bayesian optimisation starts after the 30$^{\textrm{th}}$ evaluation. The horizontal dashed line corresponds to the minimum error found using the more exhaustive MCMC search of the parameter space with the emulator, which is the target MAE for this exercise. We ran many trials of the optimisation, as indicated by the shading, each time with a different random seed. We find that the EI algorithm can locate an error close to the approximate global minimum within just 100 to 150 full model evaluations, with some runs converging on the minimum value in as few as 75 evaluations. We see very little improvement beyond 150 evaluations of the emulator. Thus, using Bayesian optimisation allows us to find a best-fitting model with roughly one thousand times fewer evaluations than MCMC (recall that the MCMC approach used 200 000 emulator evaluations).  

Although the emulator is an approximation of  \texttt{GALFORM}, it still represents a complex 15-dimensional function with properties similar to the full model, and the encouraging convergence of the EI algorithm allows us to apply the methodology to \texttt{GALFORM} with confidence.

\subsection{Applying Bayesian optimisation to \texttt{GALFORM}}

To apply Bayesian optimisation to \texttt{GALFORM}, we first evaluate the model $n$ times at points chosen in the parameter space using Sobol sampling (see \citealt{Oleskiewicz2020}), where $n$ is equal to twice the number of parameters we are investigating. In our case, since we are varying 15 model parameters, we draw an initial sample of 30 runs across the parameter space. Once these 30 \texttt{GALFORM} runs are in place we begin the optimisation process by applying the  EI algorithm as described above to decide the location in parameter space at which to run new model calculations. 
We look to the experiment in the last subsection for guidance as to when to stop the optimisation, and allow the runs to progress until at least 150 full model evaluations have been carried out. Once 150 runs have been completed, we decide to stop the optimisation if the MAE has not improved significantly compared to the MAE obtained from the best evaluation over the previous 25 steps.

\begin{figure*}
    \centering
    \includegraphics[width=1.2\textwidth,trim={1cm 4cm 0cm 2cm},clip]{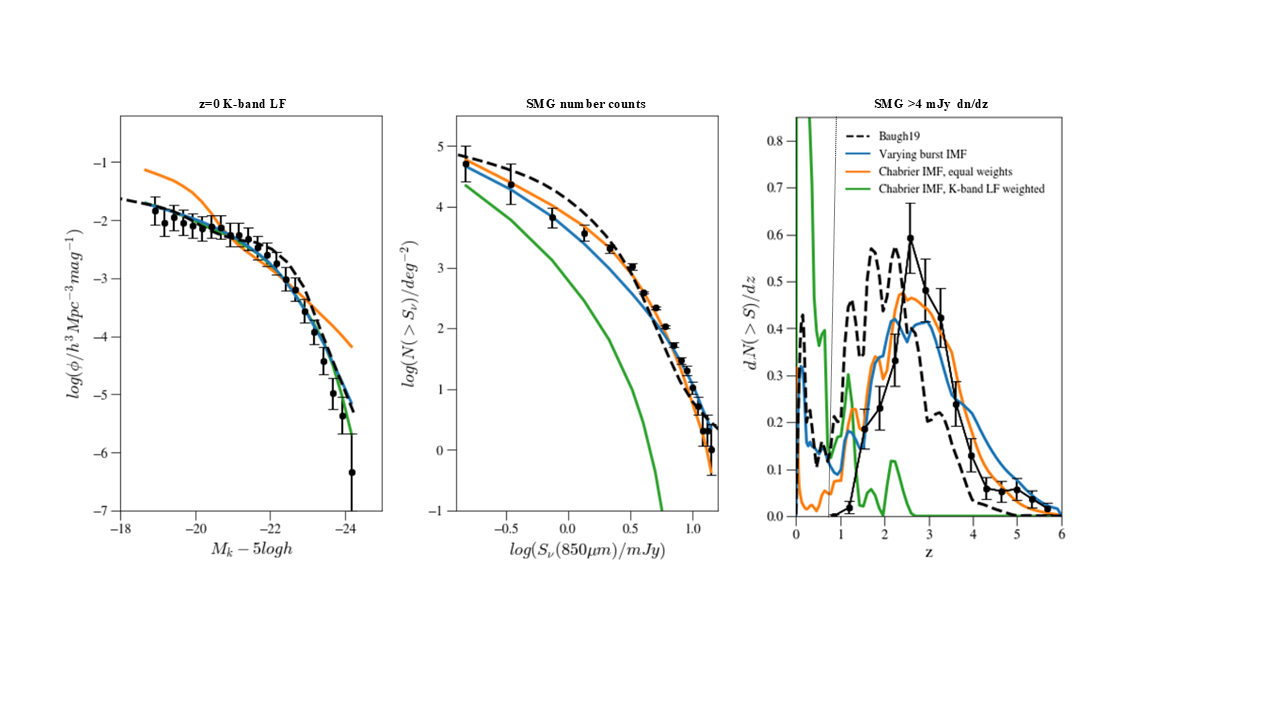}

    \caption{A comparison of the model predictions with the three calibration datasets under consideration (the parameters of these models are given in Table~2). Left: The $z = 0$ $K$-band LF. Center: the SMG number counts at $870~\mu$m. Right: the normalized SMG redshift distribution for $S_{870}>4$~mJy. The spikes in the model predictions for the redshift distribution are artefacts due to the limited number of halos simulated.
    In each case, the black points with error bars show the observational data. For the SMG redshift distribution, we calibrate to data from \protect\cite{Ugne2020}, using $z>0.8$ as indicated by the vertical dotted grey line. For the local $K$-band LF, we calibrate to data from \protect\cite{Kochanek2001}, and for the SMG number counts, we calibrate to data from \protect\cite{Stach2018} at the bright end, and \protect\cite{Chen2013} at the faint end. The orange solid curves show the model which assumes a universal Chabrier IMF in all modes of star formation. The green lines show the predictions from a model that also adopts a  universal Chabrier IMF, but which is calibrated to give an improved fit to the low-redshift $K$-band LF by increasing the weight given to this dataset in the parameter optimisation. The blue lines show a model in which the IMF slope in bursts is allowed to vary according to $ {\mathrm{d}}n/{\mathrm{d}}lnm \propto m^{-x}$, where $x$ is an adjustable parameter. For reference, the black dashed line shows the \texttt{GALFORM} model from  \protect\cite{Baugh2019}: this model was calibrated using an earlier measurement of the SMG redshift distribution from \protect\cite{Wardlow2011}, which has a lower median redshift than the \protect\cite{Ugne2020} data.}   \label{fig:main_result}
\end{figure*}

\begin{figure}
    \centering
    \includegraphics[width=0.93\columnwidth]{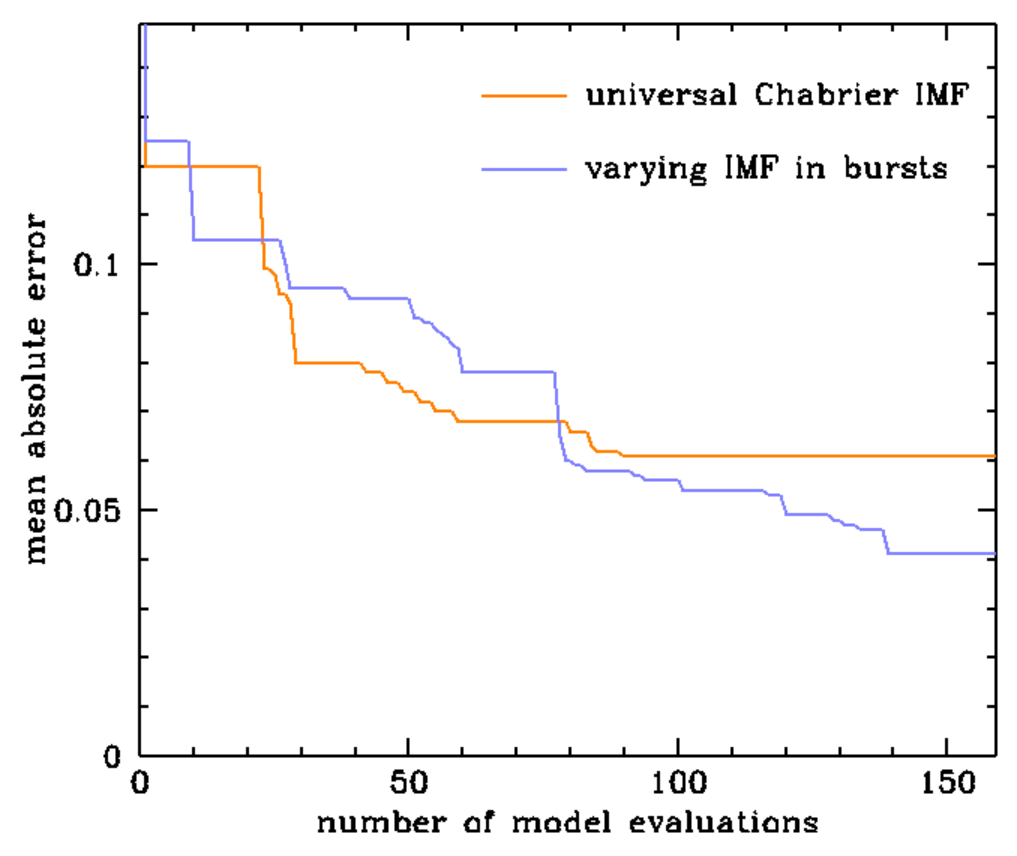}
    \caption{The minimum mean absolute error, MAE, of the \texttt{GALFORM} predictions as a function of the number of full model evaluations carried out, for the three calibration datasets: the $z=0$ $K$-band LF, the SMG number counts, and the SMG redshift distribution. The blue line shows the universal IMF model and the orange line shows the variable IMF model, in which the slope of the IMF in bursts is also a parameter; this variant gives a better (smaller) MAE. The optimisation is terminated after 165 model runs are reached and there is no significant improvement in the MAE over the preceding 25 runs.}
    \label{fig:eval}
\end{figure}

\section{Results}

Here we present the main results of calibrating \texttt{GALFORM} using Bayesian optimisation, applied to different combinations of calibration datasets (the local $K$-band LF, the number counts of SMGs and the redshift distribution of bright SMGs). Throughout we consider two 
model variants, with the main difference being whether we fix the stellar IMF in bursts of star formation to be a solar neighbourhood IMF (resulting in a 14 dimensional parameter space, referred to as the universal IMF variant) or treat the power law slope of the IMF in bursts as a model parameter (giving a 15 dimension parameter space, called the dual IMF model). In \S~4.1, we present the best-fitting models for these two variants, treating all the calibration datasets equally (the corresponding parameter values are listed in Table~2). In \S~4.2 we show how the model fits change if more weight is placed on reproducing the local calibration data. \S~4.3 presents model predictions for observational datasets that were not used in the parameter calibration.

\subsection{Calibrations}

The results of the model calibration are shown in Fig~\ref{fig:main_result} for the case in which each of the three datasets is given equal weight in the MAE metric. The associated parameter values are listed in Table~2.
(Note these plots also show a special case in which the local $K$-band luminosity function is given extra weight; this case is discussed in \S~4.2.) 
The left panel of Fig.~\ref{fig:main_result} shows the \texttt{GALFORM} predictions of the SMG redshift distribution, the middle panel shows the low redshift $K$-band LF, and the right panel the SMG number counts. We also include, for reference, the predictions from the \cite{Baugh2019} model (which assumes an IMF in  starbursts with slope $x=1$), which are shown by the black dashed lines in each panel. This model has, in general, a lower median redshift for SMGs than the best-fitting models we find here, as it was calibrated to the redshift distribution from \cite{Wardlow2011}, before the redshift distribution from \cite{Ugne2020} became available. The Baugh et~al. model is a recalibration of the model from \cite{Lacey2016} following its implementation in the P-Millennium N-body simulation. This simulation has updated cosmological parameters and a superior mass resolution compared with the N-body simulation used by Lacey et~al. However, the recalibration carried out by Baugh et~al. focussed on the local $K$-band luminosity function, but not the number counts or redshift distribution of SMGs. Furthermore, the recalibration carried out by Baugh et~al. was essentially a perturbation of the Lacey et~al. fit, considering a much smaller parameter space of only three parameters, without a framework for an extensive search of the space.

We find that the model with a universal Chabrier IMF and an equal weighting of the calibration datasets is unable to match the local $K$-band LF, the SMG redshift distribution, and the SMG number counts  \textit{at the same time}. When the calibration data does not include SMG observations, the universal IMF variant can produce a good match to the $z=0$ $K$-band LF \citep{Elliott}.
However, when the calibration datasets include SMG observations, this variant returns a poor match to the observed $K$-band LF, with large excesses at the faint- and bright-ends. At $M_{K}-5 \log h = -23$, this model predicts ten times more galaxies than are observed: this difference is many times greater than the uncertainty in the observational estimate. At the faint end, the model overpredicts the number of galaxies by at least a factor of three. Interestingly, this model prefers very low values for both $f\textsubscript{burst}$ and $F\textsubscript{stab}$, as can be seen in Table~\ref{tab:example_table}. The parameter $f\textsubscript{burst}$ is defined as the mass ratio (accreted satellite galaxy over central galaxy) threshold for a burst of star formation to occur following a merger (with the universal IMF model preferring a value of $f\textsubscript{burst} \approx 0.05$), whereas $F\textsubscript{stab}$ sets the threshold for a burst of star formation caused by the galactic disk becoming dynamically unstable to bar formation. The preferred value of $F\textsubscript{stab} = 0.53$ corresponds to a model in which there are no disk instabilities (in the formulation by \citealt{Efstathiou1982}, the ratio on the left-hand side of the expression in Eqn.~\ref{eq:diskinstab} is equal to 0.61 for a self-gravitating disk; in the model only disks for which Eqn.~\ref{eq:diskinstab} exceeds this value but is less than the adopted value of $F\textsubscript{stab}$ are allowed to become unstable and experience bursts). This combination of parameters allows the model galaxies to retain larger reservoirs of gas, which are used up in starbursts following mergers rather than disk instabilities. 

Though both models prefer a higher value of $\nu\textsubscript{SF}$, the parameter which controls the rate of quiescent star formation in disks, than is suggested by observations of galactic disks in the local Universe, the value of this parameter for the universal IMF model is significantly higher ($\nu\textsubscript{SF} = 3.48\, {\rm Gyr}^{-1}$) than the value preferred in the other two models listed in Table~\ref{tab:example_table}. This value is about a factor of 8 higher than suggested by local measurements ($\nu\textsubscript{SF} = 0.43\, {\rm Gyr}^{-1}$, with a $1-\sigma$ range of $\,0.25-0.74 \,\textrm{Gyr}^{-1}$ \citealt{Bigiel2011}). This combination of parameters, with disk instabilities in effect turned off, a high rate of minor-merger driven bursts, and much stronger quiescent star formation rates, allows the model to generate the star formation necessary to match the SMG redshift distribution and counts.  However, this behaviour means that the low redshift $K$-band LF is not matched adequately, since it leads to an excess of large, bright disk galaxies which have not undergone a recent merger, and which now make up the bright end of the LF, and which have lower mass SMBH, due to the lack of SMBH growth through disk instabilities.

For the dual IMF variant, the added flexibility of varying the IMF in starbursts allows the model to match the local $K$-band LF, as well as producing a realistic SMG redshift distribution and number counts. In this model, we see more typical values for the disk instability parameter ($f\textsubscript{stab}$ = 0.75), and a smaller number of bursts triggered by minor mergers due to a higher value of  $f\textsubscript{burst} = 0.17$. The model prefers an IMF slope parameter, $x = 0.7$, which is \textit{more} top-heavy than that assumed in \cite{Lacey2016}, who adopted $x = 1$ (though those papers were considering a larger number of calibration datasets), but is less top-heavy than assumed in early models \citep{Baugh2005}. 

The optimisation of the two \texttt{GALFORM} variants in shown in Fig.~\ref{fig:eval}, in which we plot the MAE metric as a function of the number of full runs of the model carried out. A solution is found for the universal IMF variant after 90 runs of the full \texttt{GALFORM} model, after which there is no change in the value of the MAE. In the case of the dual IMF variant, 140 model runs are needed to find a best-fitting model. The minimum MAE for the dual IMF variant is almost a factor of two smaller than that for the universal IMF model, confirming that this model gives a better reproduction of the calibration data.

\subsection{Enforcing low-redshift agreement}




\begin{figure*}    
\includegraphics[width=0.97\textwidth]{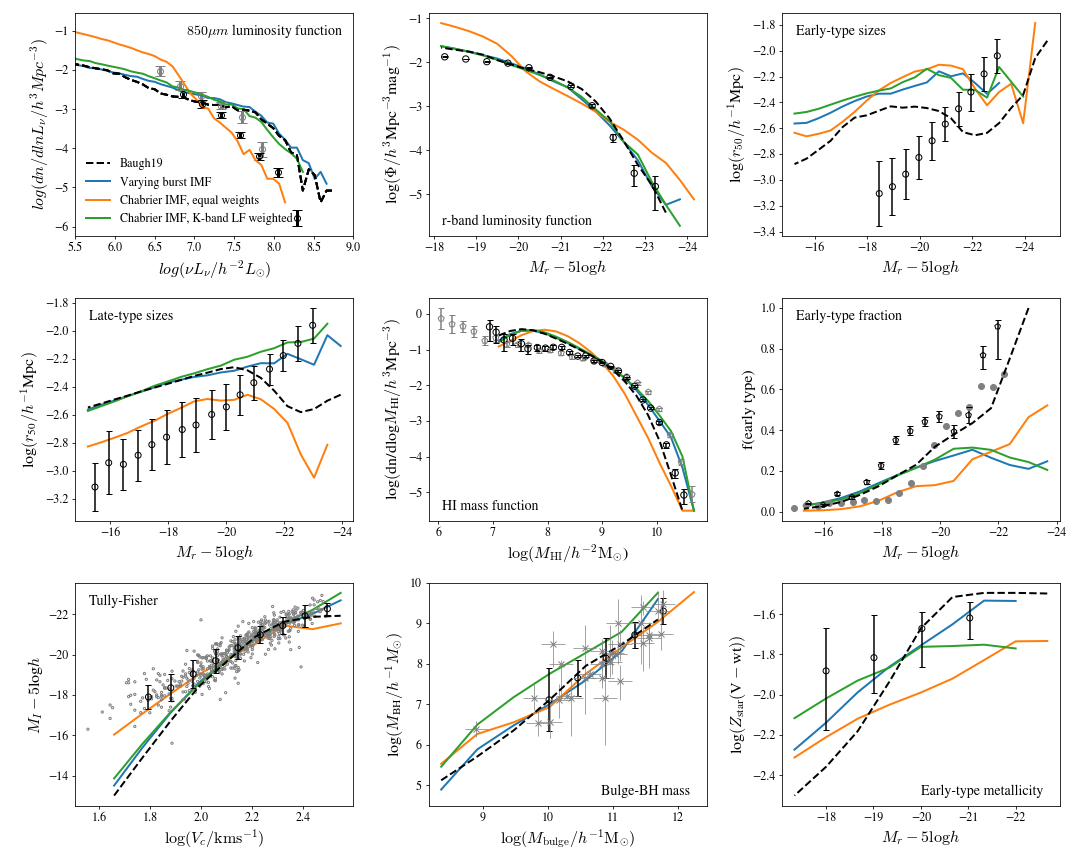}
\caption{Low redshift predictions for the three \texttt{GALFORM} calibrations: the model with a variable IMF slope in bursts (blue), the equal-weighted calibration assuming a universal Chabrier IMF (orange), and the calibration with a higher weight for the $z=0$ $K$-band LF, again assuming a universal Chabrier IMF (green). The black dashed lines show the model calibration from \protect\cite{Baugh2019}. $850\,\mu$m LF: data from \protect\cite{vlahakis2005} (grey circles) and \protect\cite{dunne2000} (black circles). $r$-band LF: data from \protect\cite{Driver2012}. Early-and late-type sizes: data from \protect\cite{Shen2003}. HI mass function: data from \protect\cite{Zwaan2005} (black circles) and \protect\cite{Martin2010} (grey circles). Early-type fraction: data derived from \protect\cite{Moffett2016} (black symbols; A. Moffett, private communication), and from \protect\cite{Gonzalez2009} (grey symbols). For the Tully-Fisher relation, we compare to a sample of Sb-Sd galaxies from the \protect\cite{Mathewson1992} catalogue selected by \protect\cite{DeJong2000} (grey points show individual galaxies, black points with bars show the binned median and 10-90 percentile range). For the bulge-BH mass relation, we compare to data from \protect\cite{Haring2004}, and for the early-type metallicity, we compare to data from \protect\cite{Smith2009}. Note that the models were not calibrated to the datasets plotted here.}
\label{fig:low_redshift_predictions}
\end{figure*}

When optimising the models in the previous subsection the three calibration datasets were weighted equally in the MAE metric. The variant with a dual IMF matches all three calibration datasets reasonably well (blue curves in Fig.~\ref{fig:main_result}). However, the calibration of the simplest variant with a universal Chabrier IMF returned a model that matched the counts and redshift distribution of SMGS but gave a poor reproduction of the local $K$-band LF (orange curves in Fig.~\ref{fig:main_result}). 

It is interesting to know how the predictions for the SMG number counts and redshift distribution are degraded if we enforce better agreement with the $z=0$ $K$-band LF. 
We can achieve this by re-calibrating the universal Chabrier IMF variant with a triple weighting applied to the low-redshift $K$-band LF in the MAE, and single weightings for each of the SMG calibration datasets. 

Running the optimisation with a higher weighting for the $z=0$ $K$-band LF, we find that the model no longer produces a realistic SMG redshift distribution or counts while simultaneously matching the $K$-band LF constraint. The green curves in Fig.~\ref{fig:main_result} show this case. Despite a much-improved match to the $K$-band LF, as expected, the predictions for the number counts are too low at all fluxes, with the deficit ranging from a factor of three at the faint end to more than a factor of a hundred at 3 mJy and brighter; the predicted median redshift is much lower than the observed one. In this weighting case, the best-fitting model parameters are more similar to the model which assumes a different IMF slope in bursts, with $F\textsubscript{stab} \approx 0.7$ for example. 

\begin{table}
	\centering
	\caption{The best-fitting parameters for the three optimisation cases. The second column shows the parameter values for the dual IMF variant, which treated the IMF slope in bursts, $x$, as another parameter. The third column shows the best-fitting parameters for the variant with a universal Chabrier IMF with equal weighting attributed to each dataset, and the model in the fourth column also assumes a universal Chabrier IMF, but with an increased weighting applied to the low-redshift $K$-band LF. $^{*}$ indicates that this parameter was held fixed. $x$ gives the slope of the IMF above $1 M_{\rm \odot}$.}
	\label{tab:example_table}
	\begin{tabular}{lccc} 
		                 \hline
        Parameter &      & variant & \\
        
  \hline
		 & Dual IMF & Universal  & Universal  \\
  		 & (IMF slope $x$  & IMF & IMF  \\
                 & in bursts)           &     & (extra weight \\
                 &           &     & to $K$ LF )\\
		\hline
		$F\textsubscript{stab}$ &  0.746 & 0.535 & 0.77\\
		$\gamma\textsubscript{SN}$ &  3.55 & 1.48 & 3.06\\
		$\alpha\textsubscript{cool}$ &  2.99 & 3.8 & 2.90\\
		$\alpha\textsubscript{reheat}$ &  1.83 & 2.14 & 2.11\\
		$V\textsubscript{SN, disk}$ (km s$^{-1}$)&  293 & 774  & 383\\
		$V\textsubscript{SN, burst}$ (km s$^{-1}$)&  349 & 399 & 194\\
		$f\textsubscript{ellip}$ &  0.383 & 0.385 & 0.212\\
		$f\textsubscript{burst}$ &  0.166 & 0.056 & 0.261\\
		$\nu\textsubscript{SF}$ (Gyr)$^{-1}$ &  1.94 & 3.48  & 2.186\\
		$f\textsubscript{SMBH}$ &  0.014 & 0.037 & 0.025\\
		$\tau\textsubscript{*burst, min}$ (Gyr) & 0.145 & 0.157 & 0.176\\
		$f\textsubscript{cloud}$ & 0.452 & 0.404  & 0.427\\
		$t\textsubscript{esc}$ (Gyr)& 0.006 &  0.006  & 0.009\\
		$\beta\textsubscript{burst}$ & 1.53 & 1.76 & 1.50  \\
		$x$ & 0.67 & $1.35^{*}$ & $1.35^{*}$\\
		
		\hline
	\end{tabular}
\end{table}

\subsection{Further predictions at low redshift}
\label{sec:further_predictions}
Having calibrated the model to the $z=0$ $K$-band LF, the SMG numbers counts, and the SMG redshift distribution, we now explore the predictions for other low-redshift properties of the galaxy population. In particular, we explore the predictions for the $z = 0$ $850 \mu m$ and $r$-band LFs, the early- and late-type galaxy sizes, the HI mass function, the dependence of the early-type fraction on $r$-band magnitude, the $I$-band Tully-Fisher relation, the black hole (BH) -bulge mass relation, and the stellar metallicities of early-type galaxies.

The model predictions for the above datasets are shown in Fig.~\ref{fig:low_redshift_predictions}. The blue solid lines show the predictions for the dual IMF \texttt{GALFORM} variant,  which allowed a variable IMF slope in bursts, whereas the orange and green lines show the predictions for variants which assumed a universal Chabrier IMF. The orange line represents model predictions with equal weightings applied to the three calibration datasets, whereas the model shown by the green line used a higher weighting for the low-redshift $K$-band LF during calibration. Interestingly, 
we see that the $z=0$ predictions for the dual IMF model and the low-redshift $K$-band LF weighted models are very similar. As discussed in \cite{Elliott}, the low-redshift $K$-band LF is most sensitive to the choice of supernova feedback parameters ($V_{\rm SN,disk}$,$V_{\rm SN,burst}$), which are similar in the case of these two calibrations. These parameters also dominate many of the other properties, leading to very similar predictions for the $z=0$ properties. For example, the Tully-Fisher relation, and late-type sizes \citep[again as shown in][]{Elliott} are dominated by the effects of the supernova feedback  parameters, at least at lower luminosities. We see that the blue and green lines are almost identical for these datasets, where the orange line better matches the faint end of both relationships.

All of the models fail to match the early-type fraction brighter than  $M\textsubscript{r} - 5 \log h  = -20$, with the low-redshift weighted Chabrier model, and the dual IMF model predicting a constant or gently decreasing early-type fraction at the brightest magnitudes. The models are therefore not producing enough high-mass early-type galaxies. This is due to a more complex interplay of parameters, and our method does not give us the tools to investigate this thoroughly. The Baugh et al. calibration, however, can match both the early-type fraction, as well as the three calibration datasets adequately while assuming a top-heavy IMF in bursts. No calibration assuming a Chabrier IMF was able to simultaneously match the three calibration datasets. 

Here, we have chosen to calibrate our model to the SMG redshift distribution at $z > 0.8$, as \cite{McAlpine2019} did when investigating the SMG population in EAGLE \citep{Schaye2015}.  The local $850 \,{\rm \mu m}$ LF therefore offers a constraint on this population at low redshift (which, at the luminosities observed, consists mainly of quiescently star-forming galaxies). We find, interestingly, that the equal-weighted Chabrier IMF fit better matches the local $850 \, {\rm \mu m}$ LF, while producing very poor fits to the $K$- and $r$-band LFs. On the other hand, the \texttt{GALFORM} variant with a dual IMF over-predicts the bright end of the local $850 \, {\rm \mu m}$ LF.

\section{CONCLUSIONS}

We have assessed whether the \texttt{GALFORM} galaxy formation model can match observations of sub-millimetre galaxies (SMGs) at 850~$\mu$m---typically high-redshift galaxies---while simultaneously reproducing properties of local galaxies. In particular, we tested the assertion by \citet{Baugh2005} that properties of both SMG and local galaxy populations can only be matched if the IMF in starbursts is assumed to be top-heavy.

Our analysis contains several advances over previous work. We used Bayesian optimisation to thoroughly search a parameter space with many dimensions (15 dimensions when the IMF slope is allowed to vary in starbursts). This method allows comprehensive parameter space searches without requiring enormous numbers of computationally expensive \texttt{GALFORM} runs, which are particularly costly when making predictions for number counts and redshift distributions. We formally quantified how well models reproduce observational data using a metric. This replaces the old-fashioned one-parameter-at-a-time variation and ``chi-by-eye'' judgement used to conclude the need for a top-heavy IMF in bursts in our previous work \citep{Baugh2005,Lacey2016}. With Bayesian optimisation, we definitively address whether the model can reproduce both SMG and local galaxy populations with a solar neighbourhood IMF.

We used three observational datasets to calibrate model parameters: the $z=0$ $K$-band luminosity function, the SMG number counts, and the SMG redshift distribution. We attempted to reproduce these datasets using two model variants: (i) a universal IMF model imposing a solar neighbourhood IMF on all star formation, and (ii) a dual IMF model with a solar neighbourhood IMF for disk star formation and a power-law IMF in bursts, with the slope treated as a model parameter.

Within the \texttt{GALFORM} framework, even when varying more parameters than explored in previous work, the model with a universal Chabrier IMF is unable to produce acceptable matches to the SMG constraints and the local $K$-band luminosity function simultaneously. This confirms the conclusion of \citet{Baugh2005} and subsequent studies \citep[e.g.,][]{Lacey2008,Lacey2016}. Here we have demonstrated the need for a top-heavy IMF with a comprehensive, automated parameter space search. We also explored calibration to the most recent SMG redshift distribution data \citep{Ugne2020}, which has a higher median redshift than previous datasets \citep[e.g.,][]{Wardlow2011}. When assuming a universal Chabrier IMF and enforcing a good fit to the low-redshift $K$-band luminosity function, we find similar results to those obtained with EAGLE \citep{McAlpine2019}: the best-fitting \texttt{GALFORM} model severely underpredicts SMG number counts and produces almost no bright sources at high redshift ($z > 1$).

Bayesian optimisation found good fits to the data using relatively few \texttt{GALFORM} evaluations, typically 100--150. In \citet{Elliott}, around 1000 \texttt{GALFORM} runs were used in an emulator-based approach (though calibrating to more datasets). Earlier works, such as \citet{Henriques2020}, used even more full model runs; our MCMC test to assess Bayesian optimisation used around 200\,000 emulator evaluations. We conclude that Bayesian optimisation is significantly faster than other methods explored for calibrating semi-analytic galaxy formation models, converging in fewer than 200 full model evaluations.

The downside of Bayesian optimisation is reduced information about model performance across the parameter space and parameter interactions. In previous emulator-based works \citep[e.g.,][]{Henriques2009, Bower2010, Vernon2010,Elliott}, emulators built from large numbers of runs were assessed for accuracy across the entire parameter space. This allows comprehensive parameter exploration, and calibration can be easily extended to include additional datasets. In \citet{Elliott}, we explored implications of calibrating to diverse dataset combinations and tensions between them with only the overhead cost of the initial 1000 runs. We could also run full MCMC explorations of parameter ranges producing acceptable matches to observational datasets. Here, however, we do not obtain a complete sense of parameter ranges matching observational datasets, as the routine searches for the global optimum rather than inferring posterior parameter distributions.

Nevertheless, we confidently conclude that we could not find any parameter set for a model with a universal Chabrier IMF that simultaneously matched the local $K$-band luminosity function and the SMG redshift distribution and number counts. This aligns with conclusions from manual searches \citep{Baugh2005,Lacey2016}. Within \texttt{GALFORM}, including the flexibility of a top-heavy IMF in bursts allows simultaneous fits to these datasets. The best-fitting dual IMF model prefers an IMF slope parameter $x = 0.7$, more top-heavy than assumed in \citet{Lacey2016} (who adopted $x=1$, though considering more calibration datasets) but less top-heavy than early models \citep{Baugh2005}.

We also examined predictions for additional low-redshift galaxy properties not used in calibration, including the local 850~$\mu$m luminosity function, the $r$-band luminosity function, early- and late-type galaxy sizes, the HI mass function, the early-type fraction versus luminosity, the Tully-Fisher relation, the black hole--bulge mass relation, and early-type stellar metallicities. The dual IMF model and the universal IMF model with enhanced $K$-band weighting produce similar predictions for most of these properties, as they prefer similar supernova feedback parameters that dominate these observables. However, all models fail to match the early-type fraction at the brightest magnitudes, suggesting areas for future model improvement.

In future work, we will extend Bayesian optimisation to recover parameter ranges for acceptable models rather than single best-fitting models, and include more observational datasets in the calibration process to improve model constraints and uncover tensions between datasets. Our work demonstrates that comprehensive, automated parameter space searches are now feasible for semi-analytic galaxy formation models, providing robust tests of fundamental assumptions such as the universality of the stellar initial mass function.

\section*{Acknowledgements}
EJE was supported by a PhD Studentship from the Durham Centre for Doctoral Training in Data Intensive Science, funded by the Science and Technology Facilities Council (STFC, ST/P006744/1) and Durham University. CMB and CGL acknowledge support from
STFC (ST/T000244/1,ST/X001075/1). We used the DiRAC@Durham facility managed by the Institute for Computational Cosmology on behalf of the STFC DiRAC HPC Facility (www.dirac.ac.uk). The equipment was funded by BEIS capital funding via STFC capital grants ST/K00042X/1, ST/P002293/1, ST/R002371/1 and ST/S002502/1, Durham University and STFC operations grant ST/R000832/1. DiRAC is part of the National e-Infrastructure.

\section*{Data Availability}
The observational datasets, \texttt{GALFORM} outputs, parameter values, and best-fitting parameter values underlying this article may be shared on reasonable request to the corresponding author.



\bibliographystyle{mnras}
\bibliography{imf_paper} 

\begin{thebibliography}{}
\makeatletter
\relax
\def\mn@urlcharsother{\let\do\@makeother \do\$\do\&\do\#\do\^\do\_\do\%\do\~}
\def\mn@doi{\begingroup\mn@urlcharsother \@ifnextchar [ {\mn@doi@}
  {\mn@doi@[]}}
\def\mn@doi@[#1]#2{\def\@tempa{#1}\ifx\@tempa\@empty \href
  {http://dx.doi.org/#2} {doi:#2}\else \href {http://dx.doi.org/#2} {#1}\fi
  \endgroup}
\def\mn@eprint#1#2{\mn@eprint@#1:#2::\@nil}
\def\mn@eprint@arXiv#1{\href {http://arxiv.org/abs/#1} {{\tt arXiv:#1}}}
\def\mn@eprint@dblp#1{\href {http://dblp.uni-trier.de/rec/bibtex/#1.xml}
  {dblp:#1}}
\def\mn@eprint@#1:#2:#3:#4\@nil{\def\@tempa {#1}\def\@tempb {#2}\def\@tempc
  {#3}\ifx \@tempc \@empty \let \@tempc \@tempb \let \@tempb \@tempa \fi \ifx
  \@tempb \@empty \def\@tempb {arXiv}\fi \@ifundefined
  {mn@eprint@\@tempb}{\@tempb:\@tempc}{\expandafter \expandafter \csname
  mn@eprint@\@tempb\endcsname \expandafter{\@tempc}}}

\bibitem[\protect\citeauthoryear{Abadi et~al.,}{Abadi
  et~al.}{2015}]{tensorflow}
Abadi M.,  et~al., 2015, {TensorFlow}: Large-Scale Machine Learning on
  Heterogeneous Systems, \url {https://www.tensorflow.org/}

\bibitem[\protect\citeauthoryear{{Bastian}, {Covey}  \& {Meyer}}{{Bastian}
  et~al.}{2010}]{Bastian2010}
{Bastian} N.,  {Covey} K.~R.,   {Meyer} M.~R.,  2010, \mn@doi [\araa]
  {10.1146/annurev-astro-082708-101642}, \href
  {https://ui.adsabs.harvard.edu/abs/2010ARA&A..48..339B} {48, 339}

\bibitem[\protect\citeauthoryear{{Baugh}}{{Baugh}}{2006}]{Baugh2006}
{Baugh} C.~M.,  2006, \mn@doi [Reports on Progress in Physics]
  {10.1088/0034-4885/69/12/R02}, \href
  {https://ui.adsabs.harvard.edu/abs/2006RPPh...69.3101B} {69, 3101}

\bibitem[\protect\citeauthoryear{{Baugh}, {Lacey}, {Frenk}, {Granato}, {Silva},
  {Bressan}, {Benson}  \& {Cole}}{{Baugh} et~al.}{2005}]{Baugh2005}
{Baugh} C.~M.,  {Lacey} C.~G.,  {Frenk} C.~S.,  {Granato} G.~L.,  {Silva} L.,
  {Bressan} A.,  {Benson} A.~J.,   {Cole} S.,  2005, \mn@doi [\mnras]
  {10.1111/j.1365-2966.2004.08553.x}, \href
  {https://ui.adsabs.harvard.edu/abs/2005MNRAS.356.1191B} {356, 1191}

\bibitem[\protect\citeauthoryear{Baugh et~al.,}{Baugh et~al.}{2019}]{Baugh2019}
Baugh C.~M.,  et~al., 2019, \mn@doi [\mnras] {10.1093/mnras/sty3427}, 483, 4922

\bibitem[\protect\citeauthoryear{Benson}{Benson}{2018}]{Benson2018}
Benson A.,  2018, \mn@doi [Research Notes of the AAS]
  {10.3847/2515-5172/aae5f4}, 2, 188

\bibitem[\protect\citeauthoryear{Benson \& Bower}{Benson \&
  Bower}{2010a}]{Benson2010a}
Benson A.~J.,  Bower R.,  2010a, \mn@doi [\mnras]
  {10.1111/j.1365-2966.2010.16592.x}, 405, 1573

\bibitem[\protect\citeauthoryear{{Benson} \& {Bower}}{{Benson} \&
  {Bower}}{2010b}]{Benson2010b}
{Benson} A.~J.,  {Bower} R.,  2010b, \mn@doi [\mnras]
  {10.1111/j.1365-2966.2010.16592.x}, \href
  {https://ui.adsabs.harvard.edu/abs/2010MNRAS.405.1573B} {405, 1573}

\bibitem[\protect\citeauthoryear{Bigiel et~al.,}{Bigiel
  et~al.}{2011}]{Bigiel2011}
Bigiel F.,  et~al., 2011, \mn@doi [\apj Letters] {10.1088/2041-8205/730/2/L13},
  730, 1

\bibitem[\protect\citeauthoryear{{Blain}, {Chapman}, {Smail}  \&
  {Ivison}}{{Blain} et~al.}{2004}]{Blain2004}
{Blain} A.~W.,  {Chapman} S.~C.,  {Smail} I.,   {Ivison} R.,  2004, \mn@doi
  [\apj] {10.1086/422353}, \href
  {https://ui.adsabs.harvard.edu/abs/2004ApJ...611..725B} {611, 725}

\bibitem[\protect\citeauthoryear{Blitz \& Rosolowsky}{Blitz \&
  Rosolowsky}{2006}]{Blitz2006}
Blitz L.,  Rosolowsky E.,  2006, \mn@doi [\apj] {10.1086/505417}, 650, 933

\bibitem[\protect\citeauthoryear{Bower, Benson, Malbon, Helly, Frenk, Baugh,
  Cole  \& Lacey}{Bower et~al.}{2006}]{Bower2006}
Bower R.~G.,  Benson A.~J.,  Malbon R.,  Helly J.~C.,  Frenk C.~S.,  Baugh
  C.~M.,  Cole S.,   Lacey C.~G.,  2006, \mn@doi [\mnras]
  {10.1111/j.1365-2966.2006.10519.x}, 370, 645

\bibitem[\protect\citeauthoryear{Bower, Vernon, Goldstein, Benson, Lacey,
  Baugh, Cole  \& Frenk}{Bower et~al.}{2010}]{Bower2010}
Bower R.~G.,  Vernon I.,  Goldstein M.,  Benson A.~J.,  Lacey C.~G.,  Baugh
  C.~M.,  Cole S.,   Frenk C.~S.,  2010, \mn@doi [\mnras]
  {10.1111/j.1365-2966.2010.16991.x}, 407, 2017

\bibitem[\protect\citeauthoryear{{Campbell} et~al.,}{{Campbell}
  et~al.}{2015}]{Campbell2015}
{Campbell} D. J.~R.,  et~al., 2015, \mn@doi [\mnras] {10.1093/mnras/stv1315},
  \href {https://ui.adsabs.harvard.edu/abs/2015MNRAS.452..852C} {452, 852}

\bibitem[\protect\citeauthoryear{{Casey}, {Narayanan}  \& {Cooray}}{{Casey}
  et~al.}{2014}]{Casey2014}
{Casey} C.~M.,  {Narayanan} D.,   {Cooray} A.,  2014, \mn@doi [\physrep]
  {10.1016/j.physrep.2014.02.009}, \href
  {https://ui.adsabs.harvard.edu/abs/2014PhR...541...45C} {541, 45}

\bibitem[\protect\citeauthoryear{Chabrier}{Chabrier}{2003}]{Chabrier2003}
Chabrier G.,  2003, \mn@doi [\pasp] {10.1086/376392}, 115, 763

\bibitem[\protect\citeauthoryear{Chapman, Blain, Smail  \& Ivison}{Chapman
  et~al.}{2005}]{Chapman2005}
Chapman S.~C.,  Blain A.~W.,  Smail I.,   Ivison R.~J.,  2005, \mn@doi [\apj]
  {10.1086/428082}, 622, 772

\bibitem[\protect\citeauthoryear{Chen, Cowie, Barger, Casey, Lee, Sanders, Wang
   \& Williams}{Chen et~al.}{2013}]{Chen2013}
Chen C.~C.,  Cowie L.~L.,  Barger A.~J.,  Casey C.~M.,  Lee N.,  Sanders D.~B.,
   Wang W.~H.,   Williams J.~P.,  2013, \mn@doi [\apj]
  {10.1088/0004-637X/762/2/81}, 762, 1

\bibitem[\protect\citeauthoryear{{Christodoulou}, {Shlosman}  \&
  {Tohline}}{{Christodoulou} et~al.}{1995}]{christodoulou1995}
{Christodoulou} D.~M.,  {Shlosman} I.,   {Tohline} J.~E.,  1995, \mn@doi [\apj]
  {10.1086/175547}, \href
  {https://ui.adsabs.harvard.edu/abs/1995ApJ...443..551C} {443, 551}

\bibitem[\protect\citeauthoryear{{Cole}, {Lacey}, {Baugh}  \& {Frenk}}{{Cole}
  et~al.}{2000}]{Cole:2000}
{Cole} S.,  {Lacey} C.~G.,  {Baugh} C.~M.,   {Frenk} C.~S.,  2000, \mn@doi
  [\mnras] {10.1046/j.1365-8711.2000.03879.x}, \href
  {https://ui.adsabs.harvard.edu/abs/2000MNRAS.319..168C} {319, 168}

\bibitem[\protect\citeauthoryear{Conroy}{Conroy}{2013}]{Conroy2013}
Conroy C.,  2013, \mn@doi [\araa] {10.1146/annurev-astro-082812-141017}, 51,
  393

\bibitem[\protect\citeauthoryear{{Conroy} \& {van Dokkum}}{{Conroy} \& {van
  Dokkum}}{2012}]{Conroy2012}
{Conroy} C.,  {van Dokkum} P.~G.,  2012, \mn@doi [\apj]
  {10.1088/0004-637X/760/1/71}, \href
  {https://ui.adsabs.harvard.edu/abs/2012ApJ...760...71C} {760, 71}

\bibitem[\protect\citeauthoryear{{Conroy}, {Gunn}  \& {White}}{{Conroy}
  et~al.}{2009}]{Conroy2009}
{Conroy} C.,  {Gunn} J.~E.,   {White} M.,  2009, \mn@doi [\apj]
  {10.1088/0004-637X/699/1/486}, \href
  {https://ui.adsabs.harvard.edu/abs/2009ApJ...699..486C} {699, 486}

\bibitem[\protect\citeauthoryear{Cowley, Lacey, Baugh, Cole, Frenk  \&
  Lagos}{Cowley et~al.}{2019}]{Cowley2019}
Cowley W.~I.,  Lacey C.~G.,  Baugh C.~M.,  Cole S.,  Frenk C.~S.,   Lagos C.
  D.~P.,  2019, \mn@doi [\mnras] {10.1093/mnras/stz1398}, 487, 3082

\bibitem[\protect\citeauthoryear{Crain et~al.,}{Crain et~al.}{2015}]{Crain2015}
Crain R.~A.,  et~al., 2015, \mn@doi [\mnras] {10.1093/mnras/stv725}, 450, 1937

\bibitem[\protect\citeauthoryear{Croton et~al.,}{Croton
  et~al.}{2006}]{Croton2006}
Croton D.~J.,  et~al., 2006, \mn@doi [\mnras]
  {10.1111/j.1365-2966.2005.09675.x}, 365, 11

\bibitem[\protect\citeauthoryear{{Da Cunha} et~al.,}{{Da Cunha}
  et~al.}{2015}]{DaCunha2015}
{Da Cunha} E.,  et~al., 2015, \mn@doi [\apj] {10.1088/0004-637X/806/1/110}, 806

\bibitem[\protect\citeauthoryear{{Dav{\'e}}, {Angl{\'e}s-Alc{\'a}zar},
  {Narayanan}, {Li}, {Rafieferantsoa}  \& {Appleby}}{{Dav{\'e}}
  et~al.}{2019}]{Dave2019}
{Dav{\'e}} R.,  {Angl{\'e}s-Alc{\'a}zar} D.,  {Narayanan} D.,  {Li} Q.,
  {Rafieferantsoa} M.~H.,   {Appleby} S.,  2019, \mn@doi [\mnras]
  {10.1093/mnras/stz937}, \href
  {https://ui.adsabs.harvard.edu/abs/2019MNRAS.486.2827D} {486, 2827}

\bibitem[\protect\citeauthoryear{Driver et~al.,}{Driver
  et~al.}{2012}]{Driver2012}
Driver S.~P.,  et~al., 2012, \mn@doi [\mnras]
  {10.1111/j.1365-2966.2012.22036.x}, 427, 3244

\bibitem[\protect\citeauthoryear{{Dudzevi{{c}}i{{u}}t{{e}}}
  et~al.,}{{Dudzevi{{c}}i{{u}}t{{e}}} et~al.}{2020}]{Ugne2020}
{Dudzevi{{c}}i{{u}}t{{e}}} U.,  et~al., 2020, \mn@doi [\mnras]
  {10.1093/mnras/staa769}, \href
  {https://ui.adsabs.harvard.edu/abs/2020MNRAS.494.3828D} {494, 3828}

\bibitem[\protect\citeauthoryear{{Dunne}, {Eales}, {Edmunds}, {Ivison},
  {Alexander}  \& {Clements}}{{Dunne} et~al.}{2000}]{dunne2000}
{Dunne} L.,  {Eales} S.,  {Edmunds} M.,  {Ivison} R.,  {Alexander} P.,
  {Clements} D.~L.,  2000, \mn@doi [\mnras] {10.1046/j.1365-8711.2000.03386.x},
  \href {https://ui.adsabs.harvard.edu/abs/2000MNRAS.315..115D} {315, 115}

\bibitem[\protect\citeauthoryear{{Efstathiou}, {Lake}  \&
  {Negroponte}}{{Efstathiou} et~al.}{1982}]{Efstathiou1982}
{Efstathiou} G.,  {Lake} G.,   {Negroponte} J.,  1982, \mn@doi [\mnras]
  {10.1093/mnras/199.4.1069}, \href
  {https://ui.adsabs.harvard.edu/abs/1982MNRAS.199.1069E} {199, 1069}

\bibitem[\protect\citeauthoryear{{Elliott}, {Baugh}  \& {Lacey}}{{Elliott}
  et~al.}{2021}]{Elliott}
{Elliott} E.~J.,  {Baugh} C.~M.,   {Lacey} C.~G.,  2021, \mn@doi [\mnras]
  {10.1093/mnras/stab1837}, \href
  {https://ui.adsabs.harvard.edu/abs/2021MNRAS.506.4011E} {506, 4011}

\bibitem[\protect\citeauthoryear{{Farr} \& {Mandel}}{{Farr} \&
  {Mandel}}{2018}]{Farr2018}
{Farr} W.~M.,  {Mandel} I.,  2018, \mn@doi [Science] {10.1126/science.aat6506},
  \href {https://ui.adsabs.harvard.edu/abs/2018Sci...361.6506F} {361, aat6506}

\bibitem[\protect\citeauthoryear{Ferrara, Bianchi, Cimatti  \&
  Giovanardi}{Ferrara et~al.}{1999}]{Ferrara1999}
Ferrara A.,  Bianchi S.,  Cimatti A.,   Giovanardi C.,  1999, \mn@doi [\apjs]
  {10.1086/313244}, 123, 437

\bibitem[\protect\citeauthoryear{{Frazier}}{{Frazier}}{2018}]{Frazier2018}
{Frazier} P.~I.,  2018, \mn@doi [arXiv e-prints] {10.48550/arXiv.1807.02811},
  \href {https://ui.adsabs.harvard.edu/abs/2018arXiv180702811F} {p.
  arXiv:1807.02811}

\bibitem[\protect\citeauthoryear{Garnett}{Garnett}{2023}]{garnett_bayesoptbook_2023}
Garnett R.,  2023, {Bayesian Optimization}.
Cambridge University Press

\bibitem[\protect\citeauthoryear{Geach et~al.,}{Geach et~al.}{2017}]{Geach2017}
Geach J.~E.,  et~al., 2017, \mn@doi [\mnras] {10.1093/mnras/stw2721}, 465, 1789

\bibitem[\protect\citeauthoryear{{Goldstein} \& {Wooff}}{{Goldstein} \&
  {Wooff}}{2007}]{bayeslinear}
{Goldstein} M.,  {Wooff} D.,  2007, Bayes linear statistics : theory and
  methods.
Wiley series in probability and statistics, John Wiley, Chichester, England

\bibitem[\protect\citeauthoryear{{Gonzalez-Perez}, {Lacey}, {Baugh}, {Lagos},
  {Helly}, {Campbell}  \& {Mitchell}}{{Gonzalez-Perez}
  et~al.}{2014}]{Gonzalez-Perez2014}
{Gonzalez-Perez} V.,  {Lacey} C.~G.,  {Baugh} C.~M.,  {Lagos} C.~D.~P.,
  {Helly} J.,  {Campbell} D.~J.~R.,   {Mitchell} P.~D.,  2014, \mn@doi [\mnras]
  {10.1093/mnras/stt2410}, \href
  {https://ui.adsabs.harvard.edu/abs/2014MNRAS.439..264G} {439, 264}

\bibitem[\protect\citeauthoryear{Gonz{\'{a}}lez, Lacey, Baugh, Frenk  \&
  Benson}{Gonz{\'{a}}lez et~al.}{2009}]{Gonzalez2009}
Gonz{\'{a}}lez J.~E.,  Lacey C.~G.,  Baugh C.~M.,  Frenk C.~S.,   Benson A.~J.,
   2009, \mn@doi [\mnras] {10.1111/j.1365-2966.2009.15057.x}, 397, 1254

\bibitem[\protect\citeauthoryear{{Granato}, {Lacey}, {Silva}, {Bressan},
  {Baugh}, {Cole}  \& {Frenk}}{{Granato} et~al.}{2000}]{Granato2000}
{Granato} G.~L.,  {Lacey} C.~G.,  {Silva} L.,  {Bressan} A.,  {Baugh} C.~M.,
  {Cole} S.,   {Frenk} C.~S.,  2000, \mn@doi [\apj] {10.1086/317032}, \href
  {https://ui.adsabs.harvard.edu/abs/2000ApJ...542..710G} {542, 710}

\bibitem[\protect\citeauthoryear{{Griffin}, {Lacey}, {Gonzalez-Perez}, {Lagos},
  {Baugh}  \& {Fanidakis}}{{Griffin} et~al.}{2019}]{Griffin2019}
{Griffin} A.~J.,  {Lacey} C.~G.,  {Gonzalez-Perez} V.,  {Lagos} C. d.~P.,
  {Baugh} C.~M.,   {Fanidakis} N.,  2019, \mn@doi [\mnras]
  {10.1093/mnras/stz1216}, \href
  {https://ui.adsabs.harvard.edu/abs/2019MNRAS.487..198G} {487, 198}

\bibitem[\protect\citeauthoryear{{Gunawardhana} et~al.,}{{Gunawardhana}
  et~al.}{2011}]{Gunawardhana2011}
{Gunawardhana} M.~L.~P.,  et~al., 2011, \mn@doi [\mnras]
  {10.1111/j.1365-2966.2011.18800.x}, \href
  {https://ui.adsabs.harvard.edu/abs/2011MNRAS.415.1647G} {415, 1647}

\bibitem[\protect\citeauthoryear{{Hamann} \& {Wons}}{{Hamann} \&
  {Wons}}{2022}]{Hamann2022}
{Hamann} J.,  {Wons} J.,  2022, \mn@doi [\jcap]
  {10.1088/1475-7516/2022/03/036}, \href
  {https://ui.adsabs.harvard.edu/abs/2022JCAP...03..036H} {2022, 036}

\bibitem[\protect\citeauthoryear{H{\"{a}}ring \& Rix}{H{\"{a}}ring \&
  Rix}{2004}]{Haring2004}
H{\"{a}}ring N.,  Rix H.-W.,  2004, \mn@doi [\apj] {10.1086/383567}, 604, L89

\bibitem[\protect\citeauthoryear{Hayward, Kere, Jonsson, Narayanan, Cox  \&
  Hernquist}{Hayward et~al.}{2011}]{Hayward2011}
Hayward C.~C.,  Kere D.,  Jonsson P.,  Narayanan D.,  Cox T.~J.,   Hernquist
  L.,  2011, \mn@doi [\apj] {10.1088/0004-637X/743/2/159}, 743

\bibitem[\protect\citeauthoryear{Hayward et~al.,}{Hayward
  et~al.}{2021}]{Hayward2021}
Hayward C.~C.,  et~al., 2021, \mn@doi [\mnras] {10.1093/mnras/stab246}, 502,
  2922

\bibitem[\protect\citeauthoryear{Henriques, Thomas, Oliver  \&
  Roseboom}{Henriques et~al.}{2009}]{Henriques2009}
Henriques B.~M.,  Thomas P.~A.,  Oliver S.,   Roseboom I.,  2009, \mn@doi
  [\mnras] {10.1111/j.1365-2966.2009.14730.x}, 396, 535

\bibitem[\protect\citeauthoryear{Henriques, Yates, Fu, Guo, Kauffmann,
  Srisawat, Thomas  \& White}{Henriques et~al.}{2020}]{Henriques2020}
Henriques B. M.~B.,  Yates R.~M.,  Fu J.,  Guo Q.,  Kauffmann G.,  Srisawat C.,
   Thomas P.~A.,   White S. D.~M.,  2020, \mn@doi [\mnras]
  {10.1093/mnras/stz3233}, 491, 5795

\bibitem[\protect\citeauthoryear{{Hickox} et~al.,}{{Hickox}
  et~al.}{2012}]{Hickox2012}
{Hickox} R.~C.,  et~al., 2012, \mn@doi [\mnras]
  {10.1111/j.1365-2966.2011.20303.x}, \href
  {https://ui.adsabs.harvard.edu/abs/2012MNRAS.421..284H} {421, 284}

\bibitem[\protect\citeauthoryear{{Hughes} et~al.,}{{Hughes}
  et~al.}{1998}]{Hughes1998}
{Hughes} D.~H.,  et~al., 1998, \mn@doi [\nat] {10.1038/28328}, \href
  {https://ui.adsabs.harvard.edu/abs/1998Natur.394..241H} {394, 241}

\bibitem[\protect\citeauthoryear{Kampakoglou, Trotta  \& Silk}{Kampakoglou
  et~al.}{2008}]{Kampakoglou2008}
Kampakoglou M.,  Trotta R.,   Silk J.,  2008, \mn@doi [\mnras]
  {10.1111/j.1365-2966.2007.12747.x}, 384, 1414

\bibitem[\protect\citeauthoryear{{Kennicutt}}{{Kennicutt}}{1983}]{Kennicutt:1983}
{Kennicutt} R.~C. J.,  1983, \mn@doi [\apj] {10.1086/161261}, \href
  {https://ui.adsabs.harvard.edu/abs/1983ApJ...272...54K} {272, 54}

\bibitem[\protect\citeauthoryear{Kochanek et~al.,}{Kochanek
  et~al.}{2001}]{Kochanek2001}
Kochanek C.~S.,  et~al., 2001, \apj, 20, 566

\bibitem[\protect\citeauthoryear{{Lacey}, {Baugh}, {Frenk}, {Silva}, {Granato}
  \& {Bressan}}{{Lacey} et~al.}{2008}]{Lacey2008}
{Lacey} C.~G.,  {Baugh} C.~M.,  {Frenk} C.~S.,  {Silva} L.,  {Granato} G.~L.,
  {Bressan} A.,  2008, \mn@doi [\mnras] {10.1111/j.1365-2966.2008.12949.x},
  \href {https://ui.adsabs.harvard.edu/abs/2008MNRAS.385.1155L} {385, 1155}

\bibitem[\protect\citeauthoryear{Lacey et~al.,}{Lacey et~al.}{2016}]{Lacey2016}
Lacey C.~G.,  et~al., 2016, \mn@doi [\mnras] {10.1093/mnras/stw1888}, 462, 3854

\bibitem[\protect\citeauthoryear{{Lagos}, {Cora}  \& {Padilla}}{{Lagos}
  et~al.}{2008}]{Lagos2008}
{Lagos} C. D.~P.,  {Cora} S.~A.,   {Padilla} N.~D.,  2008, \mn@doi [\mnras]
  {10.1111/j.1365-2966.2008.13456.x}, \href
  {https://ui.adsabs.harvard.edu/abs/2008MNRAS.388..587L} {388, 587}

\bibitem[\protect\citeauthoryear{Lagos, Lacey, Baugh, Bower  \& Benson}{Lagos
  et~al.}{2011}]{Lagos2011}
Lagos C. d.~P.,  Lacey C.~G.,  Baugh C.~M.,  Bower R.~G.,   Benson A.~J.,
  2011, \mn@doi [\mnras] {10.1111/j.1365-2966.2011.19160.x}, 416, 1566

\bibitem[\protect\citeauthoryear{{Leclercq}}{{Leclercq}}{2018}]{Leclercq2018}
{Leclercq} F.,  2018, \mn@doi [\prd] {10.1103/PhysRevD.98.063511}, \href
  {https://ui.adsabs.harvard.edu/abs/2018PhRvD..98f3511L} {98, 063511}

\bibitem[\protect\citeauthoryear{Lovell, Geach, Dav{\'{e}}, Narayanan  \&
  Li}{Lovell et~al.}{2021}]{Lovell2021}
Lovell C.~C.,  Geach J.~E.,  Dav{\'{e}} R.,  Narayanan D.,   Li Q.,  2021,
  \mn@doi [\mnras] {10.1093/mnras/staa4043}, 502, 772

\bibitem[\protect\citeauthoryear{Lu, Mo, Weinberg  \& Katz}{Lu
  et~al.}{2011}]{Lu2011}
Lu Y.,  Mo H.~J.,  Weinberg M.~D.,   Katz N.,  2011, \mn@doi [\mnras]
  {10.1111/j.1365-2966.2011.19170.x}, 416, 1949

\bibitem[\protect\citeauthoryear{Lu, Mo, Katz  \& Weinberg}{Lu
  et~al.}{2012}]{Lu2012}
Lu Y.,  Mo H.~J.,  Katz N.,   Weinberg M.~D.,  2012, \mn@doi [\mnras]
  {10.1111/j.1365-2966.2012.20435.x}, 421, 1779

\bibitem[\protect\citeauthoryear{Lu, Mo, Lu, Katz  \& Weinberg}{Lu
  et~al.}{2014}]{Lu2014}
Lu Y.,  Mo H.~J.,  Lu Z.,  Katz N.,   Weinberg M.~D.,  2014, \mn@doi [\mnras]
  {10.1093/mnras/stu1200}, 443, 1252

\bibitem[\protect\citeauthoryear{{Madar}, {Baugh}  \& {Shi}}{{Madar}
  et~al.}{2024}]{Makun}
{Madar} M.~S.,  {Baugh} C.~M.,   {Shi} D.,  2024, \mn@doi [\mnras]
  {10.1093/mnras/stae2560}, \href
  {https://ui.adsabs.harvard.edu/abs/2024MNRAS.535.3324M} {535, 3324}

\bibitem[\protect\citeauthoryear{Martin, Papastergis, Giovanelli, Haynes,
  Springob  \& Stierwalt}{Martin et~al.}{2010}]{Martin2010}
Martin A.~M.,  Papastergis E.,  Giovanelli R.,  Haynes M.~P.,  Springob C.~M.,
   Stierwalt S.,  2010, \mn@doi [\apj] {10.1088/0004-637X/723/2/1359}, 723,
  1359

\bibitem[\protect\citeauthoryear{Martindale, Thomas, Henriques  \&
  Loveday}{Martindale et~al.}{2017}]{Martindale2017}
Martindale H.,  Thomas P.~A.,  Henriques B.~M.,   Loveday J.,  2017, \mn@doi
  [\mnras] {10.1093/MNRAS/STX2131}, 472, 1981

\bibitem[\protect\citeauthoryear{{Mathewson}, {Ford}  \&
  {Buchhorn}}{{Mathewson} et~al.}{1992}]{Mathewson1992}
{Mathewson} D.~S.,  {Ford} V.~L.,   {Buchhorn} M.,  1992, \mn@doi [\apjs]
  {10.1086/191700}, \href
  {https://ui.adsabs.harvard.edu/abs/1992ApJS...81..413M} {81, 413}

\bibitem[\protect\citeauthoryear{McAlpine et~al.,}{McAlpine
  et~al.}{2019}]{McAlpine2019}
McAlpine S.,  et~al., 2019, \mn@doi [\mnras] {10.1093/mnras/stz1692}, 488, 2440

\bibitem[\protect\citeauthoryear{Moffett et~al.,}{Moffett
  et~al.}{2016}]{Moffett2016}
Moffett A.~J.,  et~al., 2016, \mn@doi [\mnras] {10.1093/mnras/stv2883}, 457,
  1308

\bibitem[\protect\citeauthoryear{Nelson et~al.,}{Nelson
  et~al.}{2015}]{Nelson2015}
Nelson D.,  et~al., 2015, \mn@doi [Astronomy and Computing]
  {10.1016/j.ascom.2015.09.003}, 13, 12

\bibitem[\protect\citeauthoryear{{Ole{\'s}kiewicz} \&
  {Baugh}}{{Ole{\'s}kiewicz} \& {Baugh}}{2020}]{Oleskiewicz2020}
{Ole{\'s}kiewicz} P.,  {Baugh} C.~M.,  2020, \mn@doi [\mnras]
  {10.1093/mnras/stz3560}, \href
  {https://ui.adsabs.harvard.edu/abs/2020MNRAS.493.1827O} {493, 1827}

\bibitem[\protect\citeauthoryear{Pillepich et~al.,}{Pillepich
  et~al.}{2018}]{Pillepich2018}
Pillepich A.,  et~al., 2018, \mn@doi [\mnras] {10.1093/mnras/stx2656}, 473,
  4077

\bibitem[\protect\citeauthoryear{Rasmussen \& Williams}{Rasmussen \&
  Williams}{2006}]{GP}
Rasmussen C.~E.,  Williams C. K.~I.,  2006, Gaussian processes for machine
  learning..
Adaptive computation and machine learning, MIT Press

\bibitem[\protect\citeauthoryear{{Robert}}{{Robert}}{2015}]{MCMC}
{Robert} C.~P.,  2015, \mn@doi [arXiv e-prints] {10.48550/arXiv.1504.01896},
  \href {https://ui.adsabs.harvard.edu/abs/2015arXiv150401896R} {p.
  arXiv:1504.01896}

\bibitem[\protect\citeauthoryear{{Robitaille}}{{Robitaille}}{2011}]{Hyperion}
{Robitaille} T.~P.,  2011, \mn@doi [\aap] {10.1051/0004-6361/201117150}, \href
  {https://ui.adsabs.harvard.edu/abs/2011A&A...536A..79R} {536, A79}

\bibitem[\protect\citeauthoryear{Rodrigues, Vernon  \& Bower}{Rodrigues
  et~al.}{2017}]{Rodrigues2017}
Rodrigues L. F.~S.,  Vernon I.,   Bower R.~G.,  2017, \mn@doi [\mnras]
  {10.1093/mnras/stw3269}, 466, 2418

\bibitem[\protect\citeauthoryear{{Rogers}, {Peiris}, {Pontzen}, {Bird}, {Verde}
   \& {Font-Ribera}}{{Rogers} et~al.}{2019}]{Rogers2019}
{Rogers} K.~K.,  {Peiris} H.~V.,  {Pontzen} A.,  {Bird} S.,  {Verde} L.,
  {Font-Ribera} A.,  2019, \mn@doi [\jcap] {10.1088/1475-7516/2019/02/031},
  \href {https://ui.adsabs.harvard.edu/abs/2019JCAP...02..031R} {2019, 031}

\bibitem[\protect\citeauthoryear{{Romano}, {Matteucci}, {Zhang}, {Papadopoulos}
   \& {Ivison}}{{Romano} et~al.}{2017}]{Romano2017}
{Romano} D.,  {Matteucci} F.,  {Zhang} Z.~Y.,  {Papadopoulos} P.~P.,   {Ivison}
  R.~J.,  2017, \mn@doi [\mnras] {10.1093/mnras/stx1197}, \href
  {https://ui.adsabs.harvard.edu/abs/2017MNRAS.470..401R} {470, 401}

\bibitem[\protect\citeauthoryear{Ruiz et~al.,}{Ruiz et~al.}{2015}]{Ruiz2015}
Ruiz A.~N.,  et~al., 2015, \mn@doi [\apj] {10.1088/0004-637X/801/2/139}, 801

\bibitem[\protect\citeauthoryear{Schaye et~al.,}{Schaye
  et~al.}{2015}]{Schaye2015}
Schaye J.,  et~al., 2015, \mn@doi [\mnras] {10.1093/mnras/stu2058}, 446, 521

\bibitem[\protect\citeauthoryear{{Schneider} et~al.,}{{Schneider}
  et~al.}{2018}]{Schneider2018a}
{Schneider} F.~R.~N.,  et~al., 2018, \mn@doi [Science]
  {10.1126/science.aan0106}, \href
  {https://ui.adsabs.harvard.edu/abs/2018Sci...359...69S} {359, 69}

\bibitem[\protect\citeauthoryear{Shen, Mo, White, Blanton, Kauffmann, Voges,
  Brinkmann  \& Csabai}{Shen et~al.}{2003}]{Shen2003}
Shen S.,  Mo H.~J.,  White S.~D.,  Blanton M.~R.,  Kauffmann G.,  Voges W.,
  Brinkmann J.,   Csabai I.,  2003, \mn@doi [\mnras]
  {10.1046/j.1365-8711.2003.06740.x}, 343, 978

\bibitem[\protect\citeauthoryear{{Silva}, {Granato}, {Bressan}  \&
  {Danese}}{{Silva} et~al.}{1998}]{Silva1998}
{Silva} L.,  {Granato} G.~L.,  {Bressan} A.,   {Danese} L.,  1998, \mn@doi
  [\apj] {10.1086/306476}, \href
  {https://ui.adsabs.harvard.edu/abs/1998ApJ...509..103S} {509, 103}

\bibitem[\protect\citeauthoryear{Simha \& Cole}{Simha \&
  Cole}{2017}]{Simha2017}
Simha V.,  Cole S.,  2017, \mn@doi [\mnras] {10.1093/MNRAS/STX1942}, 472, 1392

\bibitem[\protect\citeauthoryear{{Smail}, {Ivison}  \& {Blain}}{{Smail}
  et~al.}{1997}]{Smail1997}
{Smail} I.,  {Ivison} R.~J.,   {Blain} A.~W.,  1997, \mn@doi [\apjl]
  {10.1086/311017}, \href
  {https://ui.adsabs.harvard.edu/abs/1997ApJ...490L...5S} {490, L5}

\bibitem[\protect\citeauthoryear{{Smith}}{{Smith}}{2020}]{Smith2020}
{Smith} R.~J.,  2020, \mn@doi [\araa] {10.1146/annurev-astro-032620-020217},
  \href {https://ui.adsabs.harvard.edu/abs/2020ARA&A..58..577S} {58, 577}

\bibitem[\protect\citeauthoryear{Smith, Lucey  \& Hudson}{Smith
  et~al.}{2009}]{Smith2009}
Smith R.~J.,  Lucey J.~R.,   Hudson M.~J.,  2009, \mn@doi [\mnras]
  {10.1111/j.1365-2966.2009.15580.x}, 400, 1690

\bibitem[\protect\citeauthoryear{Smith, Hayward, Jarvis  \& Simpson}{Smith
  et~al.}{2017}]{Smith2017}
Smith D.~J.,  Hayward C.~C.,  Jarvis M.~J.,   Simpson C.,  2017, \mn@doi
  [\mnras] {10.1093/MNRAS/STX1689}, 471, 2453

\bibitem[\protect\citeauthoryear{Somerville \& Dav{\'{e}}}{Somerville \&
  Dav{\'{e}}}{2015}]{Somerville2015}
Somerville R.~S.,  Dav{\'{e}} R.,  2015, \mn@doi [\araa]
  {10.1146/annurev-astro-082812-140951}, 53, 51

\bibitem[\protect\citeauthoryear{{Somerville}, {Gilmore}, {Primack}  \&
  {Dom{\'\i}nguez}}{{Somerville} et~al.}{2012}]{Somerville2012}
{Somerville} R.~S.,  {Gilmore} R.~C.,  {Primack} J.~R.,   {Dom{\'\i}nguez} A.,
  2012, \mn@doi [\mnras] {10.1111/j.1365-2966.2012.20490.x}, \href
  {https://ui.adsabs.harvard.edu/abs/2012MNRAS.423.1992S} {423, 1992}

\bibitem[\protect\citeauthoryear{{Stach} et~al.,}{{Stach}
  et~al.}{2018}]{Stach2018}
{Stach} S.~M.,  et~al., 2018, \mn@doi [\apj] {10.3847/1538-4357/aac5e5}, \href
  {https://ui.adsabs.harvard.edu/abs/2018ApJ...860..161S} {860, 161}

\bibitem[\protect\citeauthoryear{Stein}{Stein}{1987}]{Stein1987}
Stein M.,  1987, \mn@doi [Technometrics] {10.1080/00401706.1987.10488205}, 29,
  143

\bibitem[\protect\citeauthoryear{Swinbank, Smail, Chapman, Blain, Ivison  \&
  Keel}{Swinbank et~al.}{2004}]{Swinbank2004}
Swinbank A.~M.,  Smail I.,  Chapman S.~C.,  Blain A.~W.,  Ivison R.~J.,   Keel
  W.~C.,  2004, \mn@doi [\apj] {10.1086/425171}, 617, 64

\bibitem[\protect\citeauthoryear{{Takhtaganov}, {Luki{\'c}}, {M{\"u}ller}  \&
  {Morozov}}{{Takhtaganov} et~al.}{2021}]{Takhtaganov2021}
{Takhtaganov} T.,  {Luki{\'c}} Z.,  {M{\"u}ller} J.,   {Morozov} D.,  2021,
  \mn@doi [\apj] {10.3847/1538-4357/abc8ed}, \href
  {https://ui.adsabs.harvard.edu/abs/2021ApJ...906...74T} {906, 74}

\bibitem[\protect\citeauthoryear{Vernon, Goldstein  \& Bower}{Vernon
  et~al.}{2010}]{Vernon2010}
Vernon I.,  Goldstein M.,   Bower R.~G.,  2010, Bayesian analysis., 05, 619

\bibitem[\protect\citeauthoryear{{Vlahakis}, {Dunne}  \& {Eales}}{{Vlahakis}
  et~al.}{2005}]{vlahakis2005}
{Vlahakis} C.,  {Dunne} L.,   {Eales} S.,  2005, \mn@doi [\mnras]
  {10.1111/j.1365-2966.2005.09666.x}, \href
  {https://ui.adsabs.harvard.edu/abs/2005MNRAS.364.1253V} {364, 1253}

\bibitem[\protect\citeauthoryear{{Wardlow} et~al.,}{{Wardlow}
  et~al.}{2011}]{Wardlow2011}
{Wardlow} J.~L.,  et~al., 2011, \mn@doi [\mnras]
  {10.1111/j.1365-2966.2011.18795.x}, \href
  {https://ui.adsabs.harvard.edu/abs/2011MNRAS.415.1479W} {415, 1479}

\bibitem[\protect\citeauthoryear{{Weidner}, {Ferreras}, {Vazdekis}  \& {La
  Barbera}}{{Weidner} et~al.}{2013}]{Weidner2013}
{Weidner} C.,  {Ferreras} I.,  {Vazdekis} A.,   {La Barbera} F.,  2013, \mn@doi
  [\mnras] {10.1093/mnras/stt1445}, \href
  {https://ui.adsabs.harvard.edu/abs/2013MNRAS.435.2274W} {435, 2274}

\bibitem[\protect\citeauthoryear{{Zhang}, {Romano}, {Ivison}, {Papadopoulos}
  \& {Matteucci}}{{Zhang} et~al.}{2018}]{Zhang2018}
{Zhang} Z.-Y.,  {Romano} D.,  {Ivison} R.~J.,  {Papadopoulos} P.~P.,
  {Matteucci} F.,  2018, \mn@doi [\nat] {10.1038/s41586-018-0196-x}, \href
  {https://ui.adsabs.harvard.edu/abs/2018Natur.558..260Z} {558, 260}

\bibitem[\protect\citeauthoryear{Zwaan, Meyer, Staveley-Smith  \&
  Webster}{Zwaan et~al.}{2005}]{Zwaan2005}
Zwaan M.~A.,  Meyer M.~J.,  Staveley-Smith L.,   Webster R.~L.,  2005, \mn@doi
  [\mnras: Letters] {10.1111/j.1745-3933.2005.00029.x}, 359, 1

\bibitem[\protect\citeauthoryear{de Jong \& Lacey}{de~Jong \&
  Lacey}{2000}]{DeJong2000}
de Jong R.~S.,  Lacey C.,  2000, \mn@doi [\apj] {10.1086/317840}, 545, 781

\bibitem[\protect\citeauthoryear{{van der Velden}, {Duffy}, {Croton}, {Mutch}
  \& {Sinha}}{{van der Velden} et~al.}{2019}]{vanderVelden2019}
{van der Velden} E.,  {Duffy} A.~R.,  {Croton} D.,  {Mutch} S.~J.,   {Sinha}
  M.,  2019, \mn@doi [\apjs] {10.3847/1538-4365/ab1f7d}, \href
  {https://ui.adsabs.harvard.edu/abs/2019ApJS..242...22V} {242, 22}

\bibitem[\protect\citeauthoryear{{van der Velden}, {Duffy}, {Croton}  \&
  {Mutch}}{{van der Velden} et~al.}{2021}]{vanderVelden2021}
{van der Velden} E.,  {Duffy} A.~R.,  {Croton} D.,   {Mutch} S.~J.,  2021,
  \mn@doi [\apjs] {10.3847/1538-4365/abddba}, \href
  {https://ui.adsabs.harvard.edu/abs/2021ApJS..253...50V} {253, 50}

\makeatother
\end{thebibliography}





\bsp	
\label{lastpage}
\end{document}